\documentclass[prd,reprint,superscriptaddress,floatfix,nofootinbib]{revtex4-2}

\usepackage{amsmath,amssymb,amsfonts,amsthm}
\usepackage{graphicx}
\usepackage{enumitem}
\usepackage{bm}
\usepackage{rotating}
\usepackage{hyperref}
\usepackage{pbox}
\usepackage{array}
\usepackage{mathtools}
\usepackage{leftidx}
\usepackage{lmodern}
\usepackage[normalem]{ulem}
\usepackage{braket}
\usepackage{tensor}
\usepackage[usenames,dvipsnames]{color}
\usepackage{float}
\usepackage{footnote}
\usepackage{setspace}
\usepackage{CJKutf8}

\newcommand{\Z}{\mathbb{Z}}
\newcommand{\R}{\mathbb{R}}
\renewcommand{\d}{\mathrm{d}}
\newcommand{\rr}[1]{\left(#1\right)}

\newcommand{\bc}{\text{BC}_\omega}

\newcommand{\sx}{\mathsf{x}}

\newcommand{\ii}{\text{i}}
\newcommand{\W}{\text{W}}
\newcommand{\dd}{\mathrm{d}}

\newcommand{\osc}{\text{osc}}
\newcommand{\zm}{\text{zm}}
\renewcommand{\bar}{\overline}

\begin{document}
 
\title{The time traveler's guide to the quantization of zero modes}

\author{Ana Alonso-Serrano}
\email{ana.alonso.serrano@aei.mpg.de}
\affiliation{Max-Planck-Institut f\"ur Gravitationsphysik 	(Albert-Einstein-Institut), Am M\"uhlenberg 1, D-14476 Golm, Germany}

\author{Erickson Tjoa}
\email{erickson.tjoa@uwaterloo.ca}
\affiliation{Department of Physics and Astronomy, University of Waterloo, Waterloo, Ontario, N2L 3G1, Canada}
\affiliation{Institute for Quantum Computing, University of Waterloo, Waterloo, Ontario, N2L 3G1, Canada}

\author{Luis J. Garay}
\email{luisj.garay@ucm.es}
\affiliation{Departamento de F\'isica Te\'orica and IPARCOS, Universidad Complutense de
Madrid, 28040 Madrid, Spain}
\affiliation{Instituto de Estructura de la Materia (IEM-CSIC), Serrano 121, 28006 Madrid, Spain}

\author{Eduardo~Mart\'in-Mart\'inez}
\email{emartinmartinez@uwaterloo.ca}
\affiliation{Department of Applied Mathematics, University of Waterloo, Waterloo, Ontario, N2L 3G1, Canada}
\affiliation{Institute for Quantum Computing, University of Waterloo, Waterloo, Ontario, N2L 3G1, Canada}
\affiliation{Perimeter Institute for Theoretical Physics, 31 Caroline St N, Waterloo, Ontario, N2L 2Y5, Canada}

\date{\today}

\begin{abstract}

We study the relationship between the quantization of a massless scalar field on the two-dimensional Einstein cylinder and in a spacetime with a time machine. We find that the latter picks out a unique prescription for the state of the zero mode in the Einstein cylinder. We show how this choice arises from the computation of the vacuum Wightman function and the vacuum renormalized stress-energy tensor in the time-machine geometry. Finally, we relate the previously proposed regularization of the zero mode state as a squeezed state with the time-machine warp parameter, thus demonstrating that the quantization in the latter regularizes the quantization in an Einstein cylinder.

\end{abstract}

\maketitle

\section{Motivation}

The idea of time travel has captivated imaginations and inspired science fiction for several centuries now: From the first fiction work portraying a time machine~\cite{Uribe1999} to the myriad of contemporary art inspired in wormholes~\cite{contact} and time machines~\cite{interstellar,primer,future}, the idea of time travel has fascinated many generations of both scientists and non-scientists.
From a rigorous science point of view, it is well-known that general relativity allows for solutions that have nontrivial topological or causal structures, such as G\"odel's rotating universe \cite{Godel1949solution} or wormhole spacetimes \cite{Morris1988wormhole}.  If a traversable wormhole exists, there are several ways it can be transformed in a time machine~\cite{Friedman1990cauchy}, i.e. a spacetime with closed timelike curves in part of or in the whole spacetime. In fact, it was shown that in the presence of surrounding matter, a wormhole inevitably transforms into a time machine~\cite{Novikov1990wormhole}. This relation between a wormhole and a time machine was shown to be general and not specific to a particular solution of the field equations~\cite{Frolov1991locallystatic}. It is also noteworthy that there have been theoretical proposals of self-consistent classical systems that may allow us study properties of spacetimes containing time machines \cite{Thorne1991billiard,Friedman1991Cauchy2,Ruffini2000Kinks,Politzer1994pathintegral,Friedman:1994bc}, and also proposals for experimentally feasible analogue tabletop settings resembling the properties of time-machine spacetimes~\cite{Ringbauer2014,Malpuech2021analogTM,Deutsch1991dctc}.

Due to the presence of closed timelike curves, these spacetimes generally exhibit Cauchy horizons. The stability of a viable time machine is closely related to the stability of the Cauchy horizons, which have been largely studied in the literature in the context of quantum field theory (QFT) in curved spacetimes. More concretely, the stability is investigated by analyzing divergences of the renormalized stress-energy tensor at the Cauchy horizon due to vacuum fluctuations \cite{Kim:1991mc,PhysRevD.46.3388,Frolov1991locallystatic,Kim:1991mc, Krasnikov1996stability}. One of the main difficulties in such analyses in the semiclassical regime concerns how one could define quantum field theory in these background geometries, as they  typically possess nontrivial topology and are not globally hyperbolic. {In previous studies the focus was on studying vacuum polarization effects that, due to the presence of wormholes and time machines, result in divergences of the renormalized stress-energy tensor (RSET), preventing then the ``entrance'' into the time machine.}

Quantum field theory in spacetimes with a wormhole and time machine was studied in e.g. \cite{Novikov1990wormhole,Frolov1991locallystatic}. The time-machine geometry is necessarily multiply connected, making the quantization procedure highly nontrivial. The general construction of quantum field theory on multiply-connected manifolds has been extensively studied using the framework of automorphic fields \cite{BanachDowker1979,Banach1980automorphic,Banach1979mathissues,Dowker1972multiplyconnected}. The idea is to study the same quantum field on the corresponding {universal covering space}, which has trivial topology, with certain \textit{automorphic} conditions applied to the field. A well-known example is furnished by a scalar field on the Einstein cylinder, with topology $\R\times S^1$, which in this construction is equivalent to the same scalar field in Minkowski space (the universal covering of the cylinder) with (anti-)periodic boundary conditions applied along the spatial direction.

It is known that quantization of  a massless scalar field on a topologically closed spacetime and under certain boundary conditions can give rise to \textit{zero modes}. A zero mode naturally arises when a massless scalar field is subject to periodic or Neumann boundary conditions, or when the background spacetime has toroidal topology in all spatial directions  \cite{EMM2014zeromode,tjoa2019zeroresponse,tjoa2020zeroharvest,Page_2012,Toussaint2021:2102.04284v1}. Zero modes are problematic because they do not admit a Fock representation, thus the physical ground state of the zero mode, and hence the full theory, is \textit{a priori} ambiguous \cite{EMM2014zeromode,tjoa2019zeroresponse}. For this reason zero modes are sometimes removed by hand \cite{Lin2016entangleCylin,Robles2017thermometryQFT,Brenna2016antiUnruh,braun2005entanglement,Lorek2014tripartite}, but such procedure leads to  unacceptable causality violations and other issues \cite{tjoa2019zeroresponse,tjoa2020zeroharvest}.  Some reasonable regularization of the unphysical ground state of the zero mode, using squeezed vacuum of a quantum harmonic oscillator, has been proposed in \cite{EMM2014zeromode,tjoa2019zeroresponse,tjoa2020zeroharvest} {and how the choice of regularization impacts the dynamics was studied. However, from a fundamental perspective these regularizations are essentially \textit{ad hoc} in nature. }

The attempt to regularize the zero mode naturally raises the question whether there exists a family of QFTs on curved spacetimes that can serve as the regulator. That is, we want to find a family of QFTs that do not suffer from the zero-mode problem and at the same time smoothly connect to the Einstein cylinder (e.g. as a one-parameter family). This task is nontrivial for two reasons. First, \textit{twisted} massless scalar fields on Einstein cylinder, equivalent to scalar fields with anti-periodic boundary condition $\phi(t,x+L)=-\phi(t,x)$, has no zero mode. Second, it is not obvious whether untwisted scalar fields on other spacetimes such as Misner spacetime \cite{hawking1973largescale}, despite having the same global topology $\R\times S^1$ as Einstein cylinder, exhibits zero mode (see, e.g., \cite{Gott1998Misner} and more recently \cite{emparan2021holography}). It turns out that the time-machine geometry studied in \cite{Novikov1990wormhole,Frolov1991locallystatic} provides such a one-parameter family of scalar field theories. The field theory on the Einstein cylinder is recovered in the limit of zero local curvature.

In this paper, we study the quantization of a massless scalar field in a spacetime with a time machine and analyze how a zero mode appears in the limit in which we remove the time machine. We focus on (1+1)-dimensional settings where powerful conformal techniques can be employed to obtain explicit expressions for the vacuum Wightman functions of the field. In this particular model, the time-machine spacetime can be understood as an AdS$_2$ spacetime with suitable boundary conditions applied to the field. We obtain two main results pertaining the zero mode. First, we show that while the background Einstein cylinder is obtained as the limit when the time machine disappears, the underlying quantum field theory does not smoothly approach the Einstein cylinder case because the Hadamard function (vacuum expectation of the field's anti-commutator) associated with the zero mode diverges in this limit. The Pauli-Jordan function  (vacuum expectation of the field commutator), however, \textit{does} possess a well-defined Einstein cylinder limit which \textit{includes} the zero mode contribution as required by relativistic causality \cite{tjoa2019zeroresponse}. Second, we show that the regularization of the zero-mode state first proposed in \cite{EMM2014zeromode} in terms of the squeezed vacuum of the quantum harmonic oscillator can be understood in terms of the {\it warp parameter} of the time machine. In this sense, the quantization on time-machine background prescribes a regularization for the zero-mode state in the Einstein cylinder quantization.

The paper is organized as follows. In Section~\ref{sec: geometry-spacetime} we describe a model of spacetime in (1+1) dimensions with a time machine, previously studied in \cite{Frolov1991locallystatic,Novikov1990wormhole}. In Section~\ref{sec: qft} we consider quantum field theories in the Einstein cylinder (without a time machine) and in the time-machine model. We first review the quantization of a massless scalar field in a (1+1) Einstein cylinder spacetime and then develop the quantization in the corresponding time-machine model. In Section~\ref{sec: two-point-functions} we construct the vacuum Wightman two-point functions of the massless scalar field on these background spacetimes, carefully analyzing the role of the zero mode in appropriate limits. Finally, in Section~\ref{sec: RSET}, we compute the renormalized stress-energy tensor and track the zero-mode contribution as we take the Einstein cylinder limit.

We adopt the convention that $c=\hbar=1$ and we denote by $\sx$ a spacetime point without specifying the coordinate system. The metric signature is chosen so that for a timelike vector $v^\mu$ we have $v^\mu v_\mu<0$.

\section{Geometry of wormholes and time machines}
\label{sec: geometry-spacetime}

In this section, we review the construction of a $(1+1)$ wormhole and its conversion into a time machine. In two-dimensional spacetimes, conformal techniques can be used to find closed-form expressions for various observables of interest and provide clarity to the physics at hand. Our goal is to provide enough geometrical and topological background to later study quantum field theory  on a spacetime with time machines and the properties of the vacuum state of the field.

The outline of the construction goes as follows. We will construct a $(1+1)$-dimensional asymptotically flat spacetime with a wormhole, that can then be used to produce a time machine. In two dimensions, this construction is straightforward: It amounts to topologically identifying the opposite ends of a strip that would correspond to the wormhole mouths joined by a throat. This is equivalent to an Einstein cylinder from Minkowski space (see, e.g., \cite{EMM2014zeromode,tjoa2019zeroresponse,tjoa2020zeroharvest}). If the endpoints that are identified are at different times, the construction would produce a time machine and the resulting spacetime is not isometric to the Einstein cylinder. 

Note that in higher dimensions, this procedure is developed in an analogous way by starting from a $(d+1)$-dimensional Minkowski spacetime and choosing two worldlines along which the mouths of the wormhole move. For every point on the worldlines, we can define a unique $d$-dimensional spacelike hypersurface orthogonal to the wordlines. We then proceed to cut two balls (with the same radius) in the hypersurface centered at each worldline and topologically identify the two regions where the two balls are carved out. This would correspond to introducing two wormhole {\it mouths} connecting two (possibly distant) spacetime regions, producing an orientable wormhole with (infinitesimally) short throat \cite{Friedman1990cauchy}. We remark that a wormhole can also be turned into a time machine by introducing relative motion between the two mouths \cite{Morris1988wormhole}. The generalization to include $N$ wormholes follows by introducing $N$ pairs of wormhole mouths in a similar fashion \cite{Frolov1991locallystatic}. In all these constructions, let us note that due to the topological identification, a spacetime with a wormhole or a time machine must be multiply connected, in contrast with the simply connected Minkowski space. The multiply connectedness of the underlying spacetime is captured by topological invariants such as homology classes and fundamental groups.

Before constructing the time-machine geometry (which will be multiply connected), let us first consider the most general, globally static, simply connected $(1+1)$-dimensional auxiliary spacetime $(M,g)$, where the line element associated with the metric tensor $g=g_{\mu\nu}\dd x^\mu\otimes\dd x^\nu$ in adapted coordinates reads
\begin{align}
    \d s^2 &= -\alpha(x)^2\d t^2 + \d x^2\,,
    \label{eq: metric-1}
\end{align}
with $t,x \in \R$. Here $x$ represents the proper spatial distance and the scalar curvature  is {$R=-2\alpha''/\alpha$}. From now on we will consider cases with $R\neq 0$ since the existence of time machines requires nonzero curvature. Since $\alpha(x)$ is a nowhere-vanishing function, 
without loss of generality we take $\alpha(x)>0$ and hence it will be convenient to write it in the following form
\begin{align}
    \alpha(x)=e^{-\int_0^x \d y\, a(y)}\,,
    \label{eq: alpha}
\end{align}
for some function $a(x)$.
On  $M$, $\xi=\partial_t$ is the unique (up to normalization) global timelike hypersurface-orthogonal Killing vector field, i.e. it obeys the two conditions  
\begin{align}
    \xi_{(\mu;\nu)}  = 0\,,\qquad
    \xi_{[\mu;\nu}\xi_{\lambda]}   = 0\,.
    \label{eq: hypersurface-orthogonal-killing}
\end{align}
The first condition states that $\xi=\partial_t$ is a Killing vector and the second states that it is orthogonal to the hypersurfaces of constant $t$. 
The hypersurface-orthogonality condition  guarantees that $M$ admits a foliation $M = \R\times\Sigma$, where $\Sigma$ is a Cauchy surface for $M$ orthogonal to $\xi^\mu$. Note that hypersurface orthogonality is trivially satisfied for any vector field in $(1+1)$ dimensions since $\xi_{[\mu;\nu}\xi_{\lambda]}$ is a three-form in a two-dimensional manifold. 

It will also be convenient to introduce a velocity vector field associated with \textit{Killing observers} whose four-velocity $u^\mu$
is proportional to the timelike Killing field $\xi^\mu$, i.e.
\begin{equation}
    u^\mu \coloneqq \frac{\d x^\mu}{\d \tau} = \frac{\xi^\mu}{\sqrt{-\xi^\nu\xi_\nu}}\,,
    \label{eq: Killing-observers}
\end{equation}
which in coordinates $(t,x)$ reads $u^\mu=(\alpha^{-1},0)$. The velocity field $u^\mu$ generates a flow such that at every point there is exactly one integral curve whose tangent vector is $u^\mu$, which we will call \textit{Killing trajectories}. It  is straightforward to check by direct calculation that velocity vector $u^\mu$ and the corresponding acceleration vector $a^\mu \coloneqq \d u^\mu/\d\tau$ obey the following  relations   
\begin{align}
    -a_\mu u_\nu &= u_{\mu;\nu}\,,\hspace{0.5cm}
    a_{[\mu;\nu]} = 0 \,,
    \label{eq: velocity-constraints}
\end{align}
together with the fact that $a_\mu$ for Killing observers is a simple gradient, i.e. 
\begin{align}
    a_\mu = \frac{1}{2}\nabla_\mu\log (-\xi^\nu\xi_\nu)\,,
    \label{eq: four-accel}
\end{align}
which for metric \eqref{eq: metric-1} has coordinate representation \mbox{$a_\mu=(0,-a(x))$}. Hence $a(x)$ that appears in the exponent of Eq.~\eqref{eq: alpha} takes the role of acceleration parameter of the Killing observer.  
In fact, it can be seen that the conditions \eqref{eq: hypersurface-orthogonal-killing}, which ensure that a vector field is a hypersurface-orthogonal Killing vector, are equivalent to the conditions \eqref{eq: velocity-constraints} on the corresponding velocity field and acceleration of  the Killing observers (see Appendix~\ref{appendix: four-accel}). 

In order to construct a time-machine model, we consider a new spacetime $\bar M$ by identifying points in $M$ in the following way. Given two positive  constants $A \geq 1$ and $Q$, we establish the equivalence relation $(t,x)\sim (t',x')$ if and only if $t'/t=A$ and $x'-x=Q$ in $M$.
Then the metric at  these two points  must also be identified, which implies 
\begin{align}
    A\alpha(x+Q) =\alpha(x)\,. 
    \label{eq: alpha-automorphy}
\end{align}
In other words, the function $a(x)=-\alpha'(x)/\alpha (x)$ must be a periodic function with period $Q$, whose integral over a single period is equal to $\log A$. Hence $Q$ represents the proper separation  between wormhole mouths and the \textit{warp parameter} $A$ represents the time shift that determines the ``strength'' of the time machine. When $A>1$ the spacetime will contain closed timelike curves (CTCs), as we will see shortly. This model corresponds to a time machine that possesses both a future and a past Cauchy horizon, where the CTCs are confined to the regions beyond the horizons\footnote{When considering the creation of the time machine at $t=0$ in previous literature, there exist only a future Cauchy horizon beyond with CTCs appear.} (more on this later; see also \cite{Krasnikov:1995ys}).  When taking $A\to 1$ there is no time shift and one recovers a wormhole model: in  $(1+1)$-dimensional case this is precisely the Einstein cylinder. Let us remark that this construction is essentially different from the  somewhat more conventional time machine where the identification is of the form $(t,x)\sim (t+B, x+Q)$ for some constant $B$ (that needs to be greater than the distance between wormhole mouths and contains no Cauchy horizon), see e.g. \cite{Friedman1990cauchy} for such an example.

This new spacetime $\bar M=\R\times S^1$ is locally equivalent to $M$ but is qualitatively different as far as global features are concerned. Indeed, this spacetime is multiply connected and the simply-connected spacetime $M$ is its universal cover. Although $\bar M$ has a local Killing vector $\xi$ (the same as $M$) for each simply connected region $U\subset \bar M$, this local Killing vector cannot be extended globally throughout $\bar{M}$. 
In order to see this lack of globality, it suffices to consider the norm $e^\varphi=\sqrt{-\xi^\mu\xi_\mu}$ of the Killing vector $\xi$. Let us consider the point of coordinates $(t,x)$ that can also be equally described by coordinates $(t',x')$ such that $t'=At$ and $x'=x+Q$. In both sets of coordinates, the metric has the same form given above. If $\xi$ were globally defined then both values  at $\varphi(t,x)$ and $\varphi(t',x')$ should coincide because $\varphi$ is a scalar field. However $e^{\varphi(t,x)}=\alpha(x)$ and $e^{\varphi(t',x')}=\alpha(x')=\alpha(x)/A=e^{\varphi(t,x)}/A$, hence the norm agrees only when $A=1$. Alternatively, we can also note that the local Killing field $\xi$ is defined such that its components in both coordinate charts $(t,x)$ and $(t',x')$ are given by $\xi^\mu=\xi'^\mu=(1,0)$. However, the tensor transformation law requires $\xi'^{\mu }=(\partial x'^{\mu '}/\partial x^\nu)\xi^\nu=A(1,0)=A\xi^{\mu}$, thus the components can only agree consistently throughout $\bar{M}$ when $A=1$. In other words,  $\bar M$ is locally---but not globally---static. 

In contrast, the vector fields $u^\mu$ and $a^\mu$ associated with Killing observers defined in Eq.~\eqref{eq: Killing-observers} and \eqref{eq: four-accel} can be extended globally throughout $\bar M$. The vector $u^\mu$ has components $u^\mu=(\alpha(x)^{-1},0)$ in any coordinates adapted to the local staticity. In coordinates $(t',x')$, the vector $u$ has components $u'^\mu=(\alpha(x')^{-1},0)=A(\alpha(x)^{-1},0)=Au^\mu$ and this is precisely the appropriate transformation law for a globally defined vector field under these changes of coordinates. The same applies straightforwardly to the acceleration $a_\mu$. Furthermore, as we have already seen,  they globally satisfy the conditions \eqref{eq: Killing-observers}.   

Note that the acceleration $a_\mu$ is a closed form [see Eq. (\ref{eq: velocity-constraints})]  but is  no longer an exact form, i.e. the second Killing observer condition in Eq.~\eqref{eq: hypersurface-orthogonal-killing} implies that it is locally a gradient, but not globally. Indeed, the circulation of this vector in a closed path $C$  is not zero in general. In fact, since $a_\mu=(0,-a(x))$ 
such that $x \in (0,Q)$, we find
\begin{equation}
    I[C;a]=\int_C a_\mu \dd x^\mu=\mp n\int_0^Q a(x)\dd x =\pm n I[a]\,,
    \label{eq: period-wormhole}
\end{equation}
where $n \in \Z$ (winding number) is the number of times that $C$ wraps around the wormhole and the  sign depends on the orientation of the path. 
$I[a]$ is defined as the circulation with winding number 1 and we pick an orientation providing a negative sign for concreteness. In this $(1+1)$ time machine considered here, it follows that $I[a]=-\log A$. Non-zero $I[a]$ implies that the gravitational field in a time machine is \textit{nonpotential}. Physically, it means that a particle going through a wormhole converted into a time machine can extract nonzero work since the gravitational field is not conservative. More precisely, this is because $a_\mu$ is locally exact, i.e. $a_\mu = \partial_\mu \varphi$ with $\varphi = \frac{1}{2}\log(-\xi^\nu\xi_\nu)$; however, $\varphi$ cannot be extended to a global potential (hence a {nonpotential}). Indeed, $\varphi$ has a branch cut in which it is discontinuous.
This circulation is characterized by the topology of the spacetime and characterizes the strength of the time machine. 

In order to see explicitly how the circulation characterizes the strength of the time machine, let us first note that the relation between the Killing time $t$ (such that $\xi=\partial_t$) and the  proper time $\tau$ (such that $u=\partial_\tau$) is given by   $\d \tau   = e ^{-\varphi(t,x)} \d t= \alpha(x)^{-1}\dd t $. Let us consider two points $p_1$ and $p_2$ that lie in the same $t=t_0=$ constant surface. Let $x_1$ and $x_2$ be their spatial coordinates. Consider two more  points $q_1$ and $q_2$ with the same spatial coordinates as their cousins  $x_1$ and $x_2$, both of them located in the $t=t_0+\delta t=$ constant surface. The nearby points $p_1$ and 
$q_1$ are separated by a proper time $\delta \tau_1=\alpha(x_1)^{-1}\delta t$ and likewise for the close points $p_2$ and $q_2$, i.e. $\delta \tau_2=\alpha(x_2)^{-1}\delta t$. Therefore the ratio between them is just 
$\delta \tau_2/\delta\tau_1=\alpha(x_1)/\alpha(x_2)$. If the points $p_2$, $q_2$ are obtained by translating $p_1$ and $q_1$ around the wormhole $n$ times, i.e. $x_2=x_1+nQ$, then
this ratio becomes $\delta \tau_2/\delta\tau_1=\alpha(x_1)/\alpha(x_1+nQ)=e^{-nI[a]}=A^{n}$.  

We can construct a particularly simple representative example of the time-machine geometry with warp parameter $A$ by setting the acceleration parameter $a(x)$ to be a real constant, which we define to be \mbox{$a(x) = \W \coloneqq(\log A)/L$}. We will call this representative model the \textit{canonical time machine}.
Without loss of generality we consider \mbox{$\W\geq 0$} so that $L > 0$. The canonical time-machine geometry captures all the essential topological information of the more general time-machine geometry. The only difference between the general time machine described by an arbitrary $a(x)$ of warp parameter $A$ and proper wormhole length $Q$ and the canonical one is encoded in a smooth conformal factor, which of course has no relevance in the causal structure or the topology of the time-machine spacetime.

More explicitly, any metric of the form \eqref{eq: metric-1} is conformal to that with constant $a$ equal to $\W=(\log A)/L$:
\begin{align}
    \d s^2 
    &= \Omega(y)^2(-e^{-2  \W y}\dd t^2+\dd y^2)\, ,
\end{align}
with
\begin{align}
\Omega(y) = e^{\W y}\alpha[x(y)]\, ,
\end{align}
where $x(y)$ is the solution to the separable differential equation 
\begin{align}
\frac{\d x}{\d y}= e^{  \W y}\alpha(x),
\end{align}
with the arbitrary convenient condition that the origin of $x$ and $y$ coordinates coincide.

To obtain the value of $L$ in terms of the warp parameter $A$ and the proper length $Q$ of the wormhole, we simply have to integrate  the previous equation and impose the condition that $L=y$ $(x=Q)$, i.e. that $L$ is the length of the corresponding canonical time machine. This straightforwardly gives the relation
\begin{align}
L= \frac{\log A}{A-1}\int_0^Q\frac{\d x}{\alpha(x)}.
\end{align}

From now on we will concentrate on the canonical time machine with parameters $A$ and $L$ for which the metric in coordinates $(t,y)$ in the universal covering space   $M$ is given by
\begin{equation}
    \d s^2=-e^{-2\W y}\dd t^2+\dd y^2,\,\hspace{0.5cm}\W=\frac{\log A}{L}.
    \label{eq:canonical-covering}
\end{equation}
This geometry has constant negative curvature with Ricci scalar $R=-2\W^{2}$, hence it is locally isometric to a two-dimensional Anti-de Sitter spacetime  AdS$_2$.

The canonical time-machine spacetime indeed contains CTCs as we see now and therefore so does any other conformally related to it. Indeed, consider the closed curve
\begin{align}
    y = y_0 +\frac{1}{\W}\log\frac{t}{t_0}\,,
\end{align}
where $(t_0,y_0)$ is identified with $(At_0,y_0+L)$. Its tangent vector is given by $v^\mu = (1,\W^{-1}t^{-1})$. This curve is a CTC if $v^\mu$ is timelike, i.e. if
\begin{align}
    v^\mu v_\mu <0\quad \Longleftrightarrow\quad t_0^2>\frac{1}{\W^{2}}e^{2\W x_0}\,.
    \label{eq: CTC}
\end{align}

Let us now make more explicit the connection with the Einstein cylinder. When $A\to 1$ which implies $\W\to 0$ (provided that $L$ is kept unchanged), the background geometry of the time-machine model approaches the Einstein cylinder since we have $e^{-2\W y}\to 1$ and the metric becomes flat. Conversely, this implies that if $\W > 0$, the metric is not flat, i.e. a spacetime with a time machine   is necessarily curved.  This Einstein cylinder limit is precisely what we need in order to investigate the zero-mode problem in quantum field theory in the context of time-machine model.

In order to develop a quantum field theory in next section, it is essential that $M$ can be viewed as the universal covering space for $\bar{M}$. In fact, this universal covering technique is a useful tool to work with multiply-connected spacetimes since in the universal covering space the functions are  simpler to evaluate and it is possible to define global Killing fields (if the time machine is locally static). Note that in our simple case, in order to transform results obtained in the covering space to the time-machine spacetime, we will only need to consider the role of the factor $A$ in such transformation.

For a slightly more formal review of the general machinery underlying this construction see Appendix~\ref{appendix: automorphic}.

\section{Quantum field theory}
\label{sec: qft}
 
In this section, we will describe scalar QFT on a two-dimensional spacetime with a time machine $\bar M$. We will first review the simplest QFT living on spacetime with topology $\mathbb{R}\times S^1$ where the metric is flat, commonly known as the Einstein cylinder. We will then study QFT on the topological cylinder describing a spacetime with a time machine that we have constructed in the previous section. In our discussions we will track the contribution that leads to the zero mode of the scalar field in the Einstein cylinder case. We will work with massless scalar fields since massive fields do not exhibit zero modes. 

Following Section~\ref{sec: geometry-spacetime}, we will use the  ``bar'' notation for the quantities associated with the field in the multiply-connected spacetime $\bar{M}$ (Einstein cylinder and the time machine) while the field quantities without the   ``bar'' are associated to their simply connected universal cover $M$.

\subsection{(1+1) Einstein cylinder}
\label{sec: einstein-cylinder}

Let us start with a brief recollection of the quantization of a massless scalar field $\bar{\phi}$ in a (1+1)-dimensional Einstein cylinder, with geometry locally defined by the line element
\begin{align}
    {\d s}^2 = -\d t^2+ \d y^2\,.
\end{align}
The Einstein cylinder is obtained by a topological identification given by $(t,y)\sim (t,y+L)$, where $L$ is the circumference of the cylinder and $t\in \mathbb R$. A massless scalar field $\bar{\phi}$ on the Einstein cylinder obeys the Klein-Gordon equation with periodic boundary condition
\begin{align}
   \partial_\mu\partial^\mu\bar{\phi} = 0\,,\hspace{0.5cm}\bar{\phi}(t,y) = \bar{\phi}(t,y+L)\,.
\end{align} The resulting massless quantum scalar field has Fourier mode decomposition given by \cite{EMM2014zeromode,tjoa2019zeroresponse}
\begin{align}
    \bar{\phi}(t,y) &= \bar Q_\zm(t) + \bar{\phi}_\osc(t,y)\,,\\
    \bar{\phi}_\osc(t,y) &= \sum_{n\neq 0} \frac{1}{\sqrt{4\pi|n|}}\rr{ \bar a_n^{\phantom{\dagger}} e^{-\ii|k_n|t+\ii k_n y}+ \text{h.c.}}\,,
    \label{eq: Fourier-EC}
\end{align}
where $k_n = 2\pi n/L$ and $n\in \mathbb{Z}\setminus\{0\}$. We call $\bar{\phi}_\osc$ the \textit{oscillator modes} and the spatially constant piece $\bar Q_\zm(t)$ the \textit{zero mode} because it corresponds to a zero-frequency oscillator. The ladder operators $\bar a_n^{\phantom{\dagger}},\bar a^\dagger_n$ satisfy the canonical commutation relation $[\bar a_m^{\phantom{\dagger}},\bar a_n^\dagger] = \delta_{mn}$ for all $n,m\neq 0$.

It is well known that the ground state in this theory is nontrivial because of the zero mode  \cite{EMM2014zeromode,tjoa2019zeroresponse}. We can define the Fock vacuum $\ket{\bar 0_\osc}$ for $\bar{\phi}_\osc$ as the state satisfying $\bar a_n\ket{\bar 0_\osc}=0$ for all $n\neq 0$.  However, the zero mode has no Fock representation since it is dynamically equivalent to a quantum-mechanical free particle of mass $L$. This can be seen from the observation that the   Hamiltonian is given by  
\begin{align}
    \bar H_{\zm} = \frac{\bar P_\zm^2}{2L}\,. 
\end{align}
As such, the zero mode is naturally associated with position and momentum operators $\bar Q^\textsc{s}_{\zm}, \bar P^\textsc{s}_{\zm}$ respectively (the subscript ``S'' denotes the Schr\"odinger picture). These operators satisfy equal-time canonical commutation relation $[ \bar Q^\textsc{s}_{\zm}, \bar P^\textsc{s}_{\zm}] = \ii  $. We can then express the zero mode $\bar Q_\zm(t)$ as
\begin{align}
    \bar Q_\zm(t) = \bar Q^\textsc{s}_{\zm} + \frac{\bar P^\textsc{s}_{\zm}t}{L}\,.
    \label{eq: zero-mode-operator}
\end{align}
Zero modes appear naturally in many situations, such as spacetimes with toroidal spatial topology, or when we impose Neumann boundary conditions on the field (or a mixture of Neumann and periodic boundary conditions) along all the transverse directions \cite{tjoa2019zeroresponse}. They also appear for massless scalar fields minimally coupled to curvature in de Sitter background geometry \cite{Page_2012}. 

A useful but nontrivial way of thinking about periodic boundary conditions imposed on the scalar field $\bar{\phi}$ is by considering a massless scalar field $\phi$ living on the universal cover of the Einstein cylinder (i.e. Minkowski space) where $\phi$ is subject to certain ``automorphic conditions'': the field $\bar\phi$ is then called an \textit{automorphic field} \cite{Banach1979mathissues,Dowker1972multiplyconnected,Banach1980automorphic}. These automorphic conditions are governed by the multiply-connected property of the Einstein cylinder, namely the fundamental group of the cylinder $\pi_1(\bar{M})=\mathbb{Z}$. Thus the periodicity of $\bar\phi$ comes naturally from the fact that $\bar{M}$ is a quotient space of Minkowski space associated to $\pi_1(\bar{M})$. This viewpoint will be very useful when we work with the time machine, as the scalar field properties depend explicitly on how this automorphic field construction works (see Appendix~\ref{appendix: automorphic} for an illustration and further details).

\subsection{(1+1) time-machine model}
\label{sec: standard-model}

In Section~\ref{sec: geometry-spacetime} we saw that a generic $(1+1)$ time machine $\bar M$ characterized by a warp parameter $A$ and proper wormhole length $Q$ is conformally equivalent to a canonical time-machine model characterized by the same warp parameter $A$ and proper length $L$. The canonical model differs from the generic case only in smooth local curvature contributions, since the canonical geometry has constant curvature.

In what follows we will focus on the canonical time-machine model whose universal covering space $M$ is the \textit{Poincar\'e patch} of AdS$_2$ spacetime with line element given by \eqref{eq:canonical-covering},
where $\W L=\log A$, $t\in \R$, and $y\in \R$. The canonical time-machine model $\bar M$ is the quotient space obtained by identification $(t,y)\sim (At,y+L)$ of the Poincar\'e patch, where $A\geq 1$ and $L > 0$ as we saw in  Sec.~\ref{sec: geometry-spacetime}. The fact that the universal covering space $M$ corresponds to the well-studied  Poincar\'e patch of anti-de Sitter geometry will be very helpful in understanding various properties of the quantization in the quotient space $\bar{M}$.

In order to study quantum fields on $\bar M$, we will employ universal covering techniques \cite{Dowker1972multiplyconnected, Banach1979mathissues, Banach1980automorphic,Frolov1991locallystatic}. 
These techniques allow us to generalize the quantization on the Einstein cylinder, where it reduces to imposing periodic boundary conditions on the field, to more general topological identifications. 
More concretely, the technique involves constructing the automorphic field $\bar\phi$ on the time-machine geometry from the corresponding field $\phi$ living on the Poincar\'e patch of AdS$_2$.

Starting from the metric \eqref{eq:canonical-covering}, we consider the coordinate transformation from $(t,y)$ to the more standard Poincaré-patch coordinates $(\eta,\xi)\in \mathbb R\times\mathbb R_+$ given by
\begin{align}
    \eta = t\,,\hspace{0.5cm} \xi =  e^{\W y}/\W\,.
    \label{eq:AdS-metric}
\end{align}
This transformation brings the metric into the form
\begin{align}
    \dd s^2 = \frac{1}{\W^2\xi^2}\rr{-\dd \eta^2+\dd \xi^2}\,,
    \label{eq:poincarepatch0}
\end{align}
where the AdS$_2$ length scale is given by $\W^{-1}$. In the following, it will be convenient to introduce double null coordinates in the Poincar\'e patch, defined by
\begin{align}
    \zeta_\pm = \xi \pm \eta\,.
\end{align}

\begin{figure}
    \centering
    \includegraphics{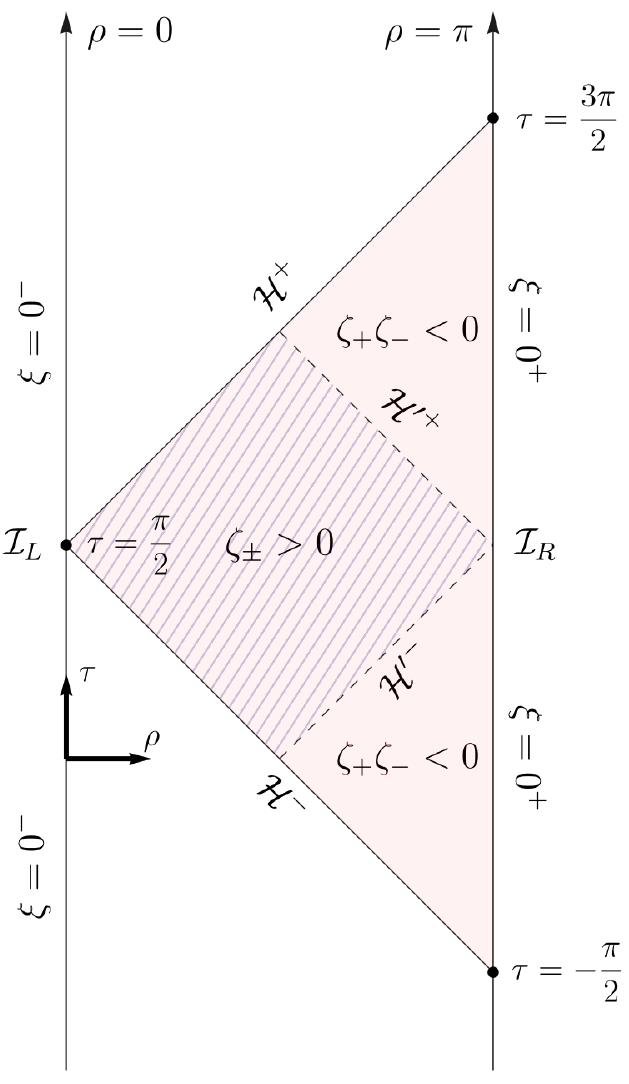}
    \caption{Conformal diagram for the Poincar\'e patch of AdS$_2$. The  Cauchy horizons  of the Poincar\'e patch are labelled $\mathcal{H}^\pm$.  After topological identification, the region $\zeta_+\zeta_-<0$ presents CTCs, thus two new Cauchy horizons $\mathcal{H'}^\pm$ appear in the conformal diagram.  In the compactified coordinates $(\tau,\rho)$, the Poincar\'e patch is covered by $\rho>|\tau-\pi/2|$ and $\rho\in (0,\pi)$.}
    \label{fig: penrose}
\end{figure}

It may be  illustrative to show the Penrose diagram of the maximal analytic extension of the Poincaré patch. With this aim we introduce the new variables $\tau,\rho$ defined by the following relations:
\begin{equation}
    \tan(\rho\pm\tau)=2\W\zeta_\pm.
\end{equation}
In these new coordinates the Poincaré patch covers the  colored region $\rho>|\tau-\pi/2|$ of the Penrose diagram of the maximal analytic extension defined by the range $\rho\in(0,\pi), \tau\in \mathbb R$ shown in Figure \ref{fig: penrose}.
The conformal boundary $\mathcal{I}$ consists of two {disconnected} pieces, namely $\mathcal{I}_{L}$ at $\rho=0$ and $\mathcal{I}_R$ at $\rho=\pi$. The Poincar\'e patch also has two past and future Cauchy horizons at $\rho=|\tau-\pi/2|$, i.e. at \mbox{$\zeta_+=\infty$}  and  $\zeta_-=\infty$, respectively, beyond which the spacetime possesses CTCs. For the time machine model, topological identification introduces new CTCs and Cauchy horizons $\mathcal{H'}^\pm$, and only the diamond-shaped region $\zeta_\pm>0$ is free of CTCs after the identification. Note that for our purposes, we apply the quantization via the universal covering approach to the standard Poincar\'e patch and not on the diamond-shaped patch, as the latter will turn out to not solve the zero mode ambiguity.

The region devoid of any CTCs after topological identification is given by $\zeta_\pm>0$ (cf. Figure~\ref{fig: penrose}) as we have already mentioned, which is consistent with Eq.~\eqref{eq: CTC}. Furthermore, recall that the standard Poincar\'e patch ($\xi>0$) itself has no CTCs in the region $\zeta_+\zeta_-<0$, but the time-machine model introduces new CTCs in these regions after topological identification.

Let us concentrate from now on  the Poincar\'e patch.
The massless Klein-Gordon equation is invariant under Weyl rescaling and reduces to the simple wave equation $(-\partial_\eta^2+\partial_\xi^2)\phi=0$ with appropriate boundary conditions at the conformal boundary $\xi=0$ as AdS$_2$ is not globally hyperbolic. (See Appendix \ref{appendix: boundary-conditions} for the choice of boundary conditions imposed on $\mathcal{I}_R$).

Since we will restrict our attention to conformally coupled massless scalar fields, we are free to choose the boundary conditions as null geodesics are able to reach the conformal boundary. For the purpose of analyzing how the zero mode arises as we vary the warp parameter $A$, it is sufficient to restrict our attention to Dirichlet boundary conditions, i.e. $\phi|_{\xi=0}=0$. We will briefly comment on the other choices of boundary conditions at the end of Section~\ref{sec: two-point-functions}.

Using the conformal invariance for a conformally coupled massless scalar field, the Klein-Gordon equation becomes (in $\zeta_\pm$ coordinates)
\begin{equation}
    \partial_+\partial_-\phi = 0\,.
\end{equation}
This has general solution $\phi(\zeta_+,\zeta_-) = F_{+}(\zeta_+)+F_-(\zeta_-)$, where $F_-$ and $F_+$ are left- and right-moving fields. On the Poincar\'e patch, a well-posed Cauchy value problem requires boundary conditions specified at the timelike boundary $\xi=0$ as we have discussed above. For Dirichlet boundary condition we have $\phi|_{\xi=0}=\phi|_{\zeta_-=-\zeta_+}=0$, hence 
the most general solution satisfying this boundary condition has the form
\begin{equation}
    \phi(\zeta_+,\zeta_-)=F(\zeta_+)-F(-\zeta_-),
\end{equation}
for some arbitrary function $F$. The Klein-Gordon product for any two solutions $\phi_1$ and $\phi_2$ can be written as
\begin{align}
    \braket{\phi_1,\phi_2} = -\ii\int_0^\infty\d\xi  \big(\phi_1\partial_\eta\phi_2^*- \phi_2^*\partial_\eta\phi_1\big)\,.
    \label{eq: KG-inner-product}
\end{align}
This inner product is independent of the choice of spacelike hypersurface $\eta=\text{constant}$.

We can obtain a set of positive frequency modes with respect to the Killing vector $\partial_\eta$ by choosing  \mbox{$F(z)=\frac1{\sqrt{4\pi\omega}}e^{-i\omega z}$}. That is, the positive-frequency eigenfunctions
\begin{align}
    u_\omega(\zeta_+,\zeta_-) = \frac{1}{\sqrt{4\pi\omega}}(e^{-\ii\omega \zeta_+}-e^{\ii\omega \zeta_-})\,,
    \label{eq: positive-frequency-eigenmode}
\end{align}
are orthonormal with respect to  the Klein-Gordon inner product \eqref{eq: KG-inner-product} in the sense that $\braket{u_\omega,u_{\omega'}}=\delta(\omega-\omega')$. Therefore any solution can be written as
\begin{equation}
    \phi(\zeta_+,\zeta_-)=\int_0^\infty  \!\!\dd \omega  \bigr[a(\omega) u_\omega(\zeta_+,\zeta_-)+a(\omega)^*u_\omega^*(\zeta_+,\zeta_-)\bigr].
\end{equation}
Promoting $a(\omega)$ and $a^*(\omega)$ to annihilation and creation operators acting on the Fock space defined in terms of the positive frequency modes above, together with commutation relation $[a(\omega) ,a ({\omega'})^\dagger]=\ii\delta(\omega-\omega')$, provides a canonical quantization of the field $\phi$.  
 
Once we have the quantum field theory defined in the covering space $M$, we would like to obtain the corresponding quantum field theory defined on the time-machine spacetime $\bar M$. We have seen in previous section the description of this spacetime, given by the periodic identification of a shifted time with parameter $A$. In terms of Poincar\'e coordinates, the time-machine geometry is obtained from the metric given by Eq.~(\ref{eq:poincarepatch0}) with the identification $(\eta,\xi)\sim A(\eta,\xi)$. We consider a fundamental domain 
given by $\W\xi\in (1,A)$ and $\eta\in \R$ and construct $\bar\phi(\zeta_+,\zeta_-)$ defined in this fundamental domain from $\phi(\zeta_+,\zeta_-)$ living on the universal cover.

In the double null coordinates, the topological identification is given by $(\zeta_+,\zeta_-)\sim A(\zeta_+,\zeta_-)$, thus the values of the (untwisted) scalar field that we are considering  will have to coincide in both identified points, i.e. $\bar\phi(A\zeta_+,A\zeta_-)=\bar\phi(\zeta_+,\zeta_-)$. Mathematically, the field $\bar\phi$ has to be automorphic under the action of the fundamental group (see Appendix~\ref{appendix: automorphic} for general definition). This automorphic requirement, which generalizes  periodic functions in Fourier theory, means that the ``annihilation variable'' $\bar a(\omega)$ must satisfy 
\begin{align}
    \bar a(\omega)=\sqrt A\ \bar a(A\omega)\,.
    \label{eq: automorphic-ladder}
\end{align}
This requirement can be satisfied by $\bar a(\omega)$ if it takes the following form
\begin{align}
    \bar a(\omega) =\omega^{-1/2} \sum_{n=-\infty}^\infty \bar c_n (\omega/\W)^{-2\pi \ii n \beta}\,,\hspace{0.25cm}\beta = \frac{1}{\log A}\,,
    \label{eq: Fourier-hidden-auto}
\end{align}
where $\bar c_n$ are arbitrary constants. In order to see this, we first express Eq.~\eqref{eq: automorphic-ladder} in terms of the function
\mbox{$f(w)\coloneqq e^{w/2}\,\bar{a}(e^w)$}, where $w=\log( \omega/\W)$.  In this language, condition~\eqref{eq: automorphic-ladder}  is simply that $f(w)$ must be periodic with period $\beta^{-1}$. Therefore, $f(w)$ can be expanded as a Fourier series
\begin{align}
    f(w)= \sum_{n=-\infty}^\infty \bar c_n e^{-2\pi \ii n \beta w}\,,
\end{align}
 showing that the condition is satisfied. 

This gives us the decomposition of the automorphic solutions $\bar\phi(\zeta_+,\zeta_-)$ as the infinite sum 
\begin{align}
    \bar\phi(\zeta_+,\zeta_-) = \sum_{n=-\infty}^\infty [\bar c_n \bar{u}_n(\zeta_+,\zeta_-)+\bar c_n^*\bar{u}^*_n(\zeta_+,\zeta_-)]\,,
\end{align}
where $\bar c_n$ are arbitrary constants and 
\begin{equation}
    \bar u_n(\zeta_+,\zeta_-)= {b_n}\int_0^\infty \frac{\dd\omega}{\omega} (\omega/\W)^{-2\pi ni\beta  }(e^{-\ii\omega\zeta_+}-e^{\ii\omega\zeta_-})\,,
    \label{eq: automorphic-modes}
\end{equation}
where $b_n$ are suitable normalization constants that will be determined below. The induced Klein-Gordon inner product in $\bar M$ evaluated on the $\eta=0$ hypersurface is simply the restriction of the Klein-Gordon product in $M$ to a fundamental domain:
\begin{align}
    (\bar\phi_1,\bar \phi_2) =-\ii\int_{\W^{-1}}^{\W^{-1} A} \d\xi\,  \big(\bar \phi_1\partial_\eta \bar \phi_2^*- \bar \phi_2^*\partial_\eta \bar \phi_1\big)\big|_{\eta=0}\,,
    \label{eq: induced-inner-prod}
\end{align}
where $\bar{\phi}_1,\bar\phi_2$ are any solutions on $\bar M$. Using this inner product \eqref{eq: induced-inner-prod} to normalize $\bar{u}_n$ (hence fixing $b_n$) and integrating over $\omega$, we obtain explicit forms for the normalized modes $\bar u_n$:
\begin{align}
    \overline{u}_0(\zeta_+,\zeta_-) &=- \rr{\frac{\beta}{4\pi}}^{\frac{1}{2}}\bigg(\ln\frac{|\zeta_+| }{|\zeta_-| }+\ii\frac{\pi}{2}(s_++s_-)\bigg)\,,
    \label{eq: u0}\\
    \overline{u}_{n\neq0}(\zeta_+,\zeta_-) &=[8\pi n \sinh(2\pi^2 \beta n)]^{-\frac{1}{2}}\notag\\
    &\hspace*{-2em}\times\bigr(e^{-\pi^2\beta n s_+}|\W\zeta_+| ^{2\pi \ii \beta n  }  -e^{\pi^2\beta n s_-}|\W\zeta_-|^{2\pi \ii \beta n }\bigr)\,,
    \label{eq: uj}
\end{align}
where $s_\pm=\text{sign}(\zeta_\pm)$. Therefore, the positive frequency modes $\bar{u}_n$ \eqref{eq: u0} and \eqref{eq: uj} form an orthonormal basis of the positive frequency one-particle Hilbert space, i.e. $(\bar u_n, \bar u_{n'})=\delta_{nn'}$. 

We close this section by making a few remarks regarding the range of $\zeta_\pm$ that we will consider in the subsequent calculations. Due to the identification $(\eta,\xi)\sim A(\eta,\xi)$, we will only perform calculations in the region where there is no CTC \textit{after topological identification}, as done in \cite{Frolov1991locallystatic}. This corresponds to the proper subset of the Poincar\'e patch given by the diamond-shaped region $\zeta_\pm>0$. Note that the field $\phi$ itself is quantized in the \textit{full} Poincar\'e patch before topological identification even though the calculations of observables (such as the two-point functions or stress-energy tensor) are restricted to the diamond-shaped region without CTCs. This is the key to extract the zero mode regularization we propose in this work. 

Another way to see this is to consider what happens if instead we try to quantize $\phi$ only in the diamond-shaped proper subset of the universal covering space. For this diamond-shaped region, there is an adapted coordinate system given by 
\begin{align}
    \eta &=\frac{1}{\W}e^\chi \tan \sigma\,,\quad
    \xi =\frac{1}{\W}e^\chi\sec\sigma\,,
    \label{eq: adapted-coordinates}
\end{align}
in terms of which the metric becoems
\begin{align}
    \d s^2=\frac{1}{\W^2}(-\d\sigma^2+\cos^2\sigma \d\chi^2),
\end{align}
with $|\sigma|<\pi/2$ and $\chi \in \mathbb{R}$. In these coordinates, the topological identification now reads $(\sigma,\chi)\sim(\sigma,\chi+\log A)$.  Notice that this is precisely the same identification as the Einstein cylinder, thus the vacuum state associated with the conformal Killing time $\sigma$ would seem to possess the same zero mode ambiguity as the Einstein cylinder. Consequently, our task cannot be achieved by canonical quantization in the diamond-shaped region.

On the other hand, in these $\chi,\sigma$ coordinates, valid only inside the diamond, one can easily see that 
\begin{align}
    \frac{\zeta_+}{\zeta_-} = \frac{1+\sin\sigma}{1-\sin\sigma}\,,
    \label{eq: log-sigma}
\end{align}
so the mode $\bar{u}_0$, given in \eqref{eq: u0}, of the quantization in the full Poincar\'e patch, when restricted to the diamond-shaped region depends only on the timelike coordinate~$\sigma$. We note that in this sense, even though $u_0$ is not a zero mode in our quantization, when restricted to the diamond $\zeta_\pm>0$ it looks like the zero mode of the quantization restricted to the diamond.

In summary, according to the automorphic prescription we obtain  $\bar{\phi}(\zeta_+,\zeta_-)$ for the time-machine geometry $\bar M$ by restricting $\phi(\zeta_+,\zeta_-)$ to take values on the fundamental domain $\bar{M}$, i.e. $\eta\in\R$ and $\xi\in (1,A)$.  The corresponding canonical quantization  is carried out by promoting the constants $\bar c_n^{\phantom{*}}$ and $\bar c_n^*$ to annihilation and creation operators acting  on the Fock space defined in terms of the positive frequency modes $\bar u_n$, with canonical commutation relations $[\bar c_n^{\phantom{\dag}},\bar c_{n'}^\dag]=\ii \delta_{nn'}$.

\section{Vacuum two-point functions 
\label{sec: two-point-functions}}

In this section, we will analyze several vacuum two-point functions and understand the limiting behaviour from the time machine to the Einstein cylinder in terms of these functions. These two-point functions are very useful for many purposes, such as detector responses upon interacting with the field within the Unruh-DeWitt model \cite{EMM2014zeromode}, relativistic causality \cite{tjoa2019zeroresponse} and communication \cite{Jonsson2020communication,Simidzija2020communication}, or computing the renormalized stress-energy tensor of the field~\cite{birrell1984quantum}.

In our model, the universal-covering approach described earlier will enable us to compute the vacuum two-point functions in terms of the orthonormal basis $\overline{u}_n$ in $\overline M$. In particular, by computing the commutator and anti-commutator vacuum expectation values, we will obtain the vacuum Wightman two-point functions for the time-machine geometry. 

The relevant vacuum correlators, namely the Wightman function $W(\sx,\sx')$, the Hadamard function (anti-commutator vacuum expectation value) $C^+(\sx,\sx')$, and the Pauli-Jordan function (commutator vacuum expectation value) $C^-(\sx,\sx')$, are given by
\begin{subequations}
\begin{align}
    W(\sx,\sx')   &\coloneqq \braket{0|\bar\phi(\sx)\bar\phi(\sx')|0}\,,\\
    C^+(\sx,\sx') &\coloneqq \braket{0|\{\bar\phi(\sx),\bar\phi(\sx')\}|0}\,,\\
    C^-(\sx,\sx') &\coloneqq \braket{0|[\bar\phi(\sx),\bar\phi(\sx')]|0}\,,
\end{align}
\end{subequations}
where $\sx$ is the shorthand for the spacetime points in any coordinates. They are related by
\begin{align}
    C^\pm(\sx,\sx') = W(\sx,\sx') \pm W(\sx',\sx)\,.
    \label{eq: Cpm-cylinder}
\end{align}

In the following, we will first review the two-point functions for the Einstein cylinder, focusing on the presence of a zero mode contribution. Then we will proceed to calculate those for the canonical time machine and determine the behavior in the limiting case where the Einstein cylinder is recovered.

\subsection{Einstein cylinder}

We have already analyzed the mode decomposition of the field in oscillatory modes and the zero mode. In terms of it, let us denote the oscillator and the zero mode vacuum two-point functions of the Einstein cylinder by $C^\pm_\osc(\sx,\sx'),W_\osc(\sx,\sx')$ and $C^\pm_\zm(\sx,\sx'),W_\zm(\sx,\sx')$ respectively. The vacuum on the Einstein cylinder can be written as the product state $\ket{0}=\ket{0_\zm}\otimes \ket{0_\osc}$. By construction we consider $\sx = (t,y)$ where $t\in \R$ and $y\in [0,L]$.

Using the mode decomposition for $\bar{\phi}_\osc$ in Eq.~\eqref{eq: Fourier-EC}, the normalized positive-frequency modes on the Einstein cylinder read
\begin{align}
\bar{u}_{n}(\sx) &= \frac{1}{\sqrt{4\pi n}}e^{-\ii |k_n| t+\ii k_n y}\,,
\end{align}
with $k_n=2\pi n/{L}$. It follows that the Wightman function is given by 
\begin{align}
    &W_\osc(\sx,\sx') 
    = \sum_{n\neq 0}\bar{u}_n(\sx)\bar{u}_n^*(\sx') \notag\\
    &= -\frac{1}{4\pi}\bigg[\log\bigg(1-e^{\frac{2\pi \ii \Delta z_-}{L}}\bigg)+\log\bigg(1-e^{-\frac{2\pi \ii \Delta z_+ }{L}}\bigg)\bigg]\,,
    \label{eq: EC-oscillator-Wightman}
\end{align}
where $z_- = y-t$, $z_+= y + t$ are the null coordinates and $\Delta z_- = z_- -z_-'$, $\Delta z_+ = z_+-z_+'$. We have dropped the $\ii\epsilon$ term for simplicity
since it can be interpreted as prescribing the branch cuts for different distributions.
Consequently, the Hadamard and Pauli-Jordan functions can be readily obtained from this expression using Eq.~\eqref{eq: Cpm-cylinder} read 
\cite{tjoa2020zeroharvest}
\begin{align}
    &C^\pm_\osc(\sx,\sx')\notag\\
    &= \mp\frac{1}{4\pi}\left[\log\bigg(1-e^{-\frac{2\pi \ii \Delta z_-}{L}}\bigg) +\log\bigg(1-e^{\frac{2\pi \ii \Delta z_+}{L}}\bigg)\right]\notag\\
    &\hspace{0.4cm}-\!\frac{1}{4\pi}\left[\log\bigg(1-e^{\frac{2\pi \ii \Delta z_-}{L}}\bigg)+\log\bigg(1-e^{-\frac{2\pi \ii \Delta z_+ }{L}}\bigg)\right]\,,
    \label{eq: osc-hadamard}
\end{align}
where as before, we have removed the $\ii\epsilon$ for simplicity.

For the zero mode, there is no \textit{a priori} good ground state because the energy eigenstate of the Hamiltonian with zero eigenvalue is not normalizable: we know that the momentum eigenstate $\ket{p}$ has Dirac delta normalization $\braket{p|p'} = \delta(p-p')$. A natural (but nonetheless \textit{ad hoc}) alternative would be to assume that a physical ground state for the zero mode can be taken to be the ground state of a quantum harmonic oscillator described by the following first and second moments \cite{EMM2014zeromode}
\begin{align}
    \braket{Q_\textsc{s}} &= \braket{P_\textsc{s}} = \braket{\{Q_\textsc{s},P_\textsc{s}\}} = 0\,,
    \label{eq: moments-1}\\
    \braket{Q^2_\textsc{s}} &= \frac{1}{2\gamma}\,,\hspace{0.5cm} \braket{P_\textsc{s}^2} = \frac{\gamma}{2}\,,
    \label{eq: moments-2}
\end{align}
where $\gamma=\sqrt{m\omega}$ is a dimensionless frequency parameter associated to the mass and natural frequency of the oscillator. This choice is natural because it includes the free-particle momentum eigenstate as a (singular) limit $\gamma\to 0$ and the usual property of a ground state of being a Gaussian state. Thus, one can think of the momentum eigenstate as the limiting case of highly squeezed vacuum state of a harmonic oscillator along the momentum direction in phase space. 

The Wightman function for the zero mode is computed using the Heisenberg operator defined in Eq.~\eqref{eq: zero-mode-operator}, together with the first and second moments \eqref{eq: moments-1}-\eqref{eq: moments-2}. We can then show that \cite{tjoa2019zeroresponse,tjoa2020zeroharvest}
\begin{align}
C^+_\zm(t,t') &= \frac{1}{\gamma} + \gamma\frac{tt'}{L^2}\,,\label{eq: zm-hadamard}\\
C^-_\zm(t,t') &= -\frac{\ii \Delta t}{L}\,,\label{eq: zm-commutator}\\
    W_\zm(t,t') &= \frac{1}{2\gamma} + \gamma\frac{tt'}{2L^2} -\frac{\ii\Delta t}{2L}\,,
\end{align}
where $\Delta t = t-t'$. The zero-mode Wightman function is translation invariant along the spatial direction but not time-translation invariant, as the second term contains the product $tt'/L^2$. Furthermore, we see that the limit $\gamma\to 0$ of $W_\zm(t,t')$ is divergent, which is equivalent to the statement that the momentum eigenstate of $\hat P_\textsc{s}$ is not a valid physical state of the zero mode.

\subsection{Time machine-model}

We will now explicitly carry the calculation of the various vacuum two-point functions for the canonical time-machine model. First note that the Wightman two-point function can be expressed in terms of sums of the mode functions $\overline u_n$ given in Eqs.~\eqref{eq: u0} and \eqref{eq: uj}: 
\begin{align}
    W(\sx,\sx')  = \sum_{n=-\infty}^\infty \overline u_n(\sx)\overline u_n^*(\sx')\,.
\end{align}
By using the relation among the two-point functions, we write the Wightman function in terms of the Hadamard function (anti-commutator) $C^+$ and the Pauli-Jordan function (commutator) $C^-$
\begin{align}
    W(\sx,\sx') &= \frac{1}{2}\left[C^+(\sx,\sx')+C^-(\sx,\sx')\right]\,,\\
    C^+ (\sx,\sx') &= \braket{0|\{\bar{\phi}(\sx),\bar{\phi}(\sx')\}|0} \notag\\
    &= \sum_{n}\left[\bar u_n(\sx)\overline u_n^*(\sx') + \overline u_n^*(\sx)\overline u_n(\sx')\right]\,,
    \label{eq: anticommutator-from-wightman}\\
    C^- (\sx,\sx') &= \braket{0|[\bar{\phi}(\sx),\bar{\phi}(\sx')]|0} \notag\\
    &= \sum_{n}\left[\overline u_n(\sx)\overline u_n^*(\sx') - \overline u_n^*(\sx)\overline u_n(\sx')\right]\,.
    \label{eq: commutator-from-wightman}
\end{align}

We derive now the expressions of  the Hadamard and Pauli-Jordan functions that will determine the Wightman function. Let us first consider the Hadamard function. Using the mode sum formulation for it, the Hadamard function reads
\begin{align}
    C^+(\sx,\sx') &= C^+_0(\sx,\sx') + C^+_1(\sx,\sx')+ C^+_2(\sx,\sx')\,,
\end{align}
where term $C^+_0$ is given by the mode functions ${\bar{u}_0}$ of Eq.~(\ref{eq: u0}) and  $C^+_1$ and $C^+_2$ from the contributions of the two different terms of the mode functions ${\bar{u}_n}$ in Eq. (\ref{eq: uj}). By computing the different terms, the exponential functions will give rise to different trigonometric contributions in $\beta$. After some tedious but straightforward algebraic manipulations, we get
\begin{align}
    &C^+_0(\sx,\sx')=\frac{\beta}{2\pi}\Bigr[\log\left|\frac{\zeta_+'}{\zeta_-'}\right|\log\left|\frac{\zeta_+}{\zeta_-}\right|
    \notag\\
    &\hspace{0.3cm}+\frac{\pi^2}{4}(s_++s_-)(s'_++s'_-)\Bigr]\,,\\
    &C^+_1(\sx,\sx')=  \frac{1}{2\pi}\sum_{n=1}^\infty \frac{1}{n\sinh (2\pi^2\beta n)}\notag\\
    &\hspace{0.3cm}\times \bigg[\cosh \Bigr(\pi^2\beta n(s_++s'_+)\Bigr) \cos\rr{2\pi\beta n \log\left|\frac{\zeta_+'}{\zeta_+}\right|}\notag\\
    &\hspace{0.5cm}+\cosh \Bigr(\pi^2\beta n(s_-+s'_-)\Bigr) \cos\rr{2\pi\beta n\log\left|\frac{\zeta_-'}{\zeta_-}\right|}\bigg]\,,\\
    &C^+_2(\sx,\sx')=- \frac{1}{2\pi}\sum_{n=1}^\infty \frac{1}{n\sinh (2\pi^2\beta n)}\notag\\
    &\hspace{0.3cm}\times \bigg[\cosh \Bigr(\pi^2\beta n(s'_--s_+)\Bigr) \cos\rr{2\pi\beta n\log\left|\frac{\zeta_+'}{\zeta_-}\right|}\notag\\
    &\hspace{0.5cm}+\cosh \Bigr(\pi^2\beta n(s_--s'_+)\Bigr) \cos\rr{2\pi\beta n\log\left|\frac{\zeta_-'}{\zeta_+}\right|}\bigg]\,.
    \label{eq: ci+}
\end{align}
As before we have dropped the $\ii\epsilon$ term for simplicity
since it can be interpreted as prescribing the branch cuts for different distributions.

For the quantization in the whole patch, we are now interested in calculating the two point functions just inside the diamond, where $\zeta_\pm> 0$ (i.e. $s_\pm=1$) and there are no CTCs. This restriction allow us to directly compare it to the Einstein cylinder and use it to select a state for the zero mode. In this particular region, we can express $C^+(\sx,\sx')$ in terms of Jacobi theta functions instead of series~\cite{abramowitz1972handbook}. The Hadamard function would then read \cite{Frolov1991locallystatic} 
\begin{align}
    &C^+(\sx,\sx') = \frac{\beta}{2\pi}\rr{\log\frac{\zeta_+'}{\zeta_-'}\log\frac{\zeta_+}{\zeta_-}} \notag\\
    &
    -\frac{1}{2\pi}\log\left(\frac{\theta_1(\beta\pi\log(\zeta'_+/\zeta_+))\theta_1(\beta\pi\log(\zeta'_-/\zeta_-))}{\theta_4(\beta\pi\log(\zeta'_+/\zeta_-))\theta_4(\beta\pi\log(\zeta'_-/\zeta_+))}\right)\,.
    \label{eq: hadamard-jacobi-frolov}
\end{align}
In this form, the shorthand $\theta_j(z)\equiv \theta_j(z,e^{-2\pi^2\beta})$ is used\footnote{Note that our expression here has an extra factor of $\pi$ in the argument of the Jacobi theta function $\theta_j(z)$, which is the standard notation in many handbooks (e.g. \cite{abramowitz1972handbook,gradshteyn2014table}) and symbolic computation software such as \textit{Mathematica} (written as \textsf{EllipticTheta} \cite{JacobiTh39:online}). The convention for $\theta_j(z)$ in \cite{Frolov1991locallystatic} \textit{without} $\pi$ is in fact $\vartheta_j(z)$ from McKean and Moll’s notation, where $\vartheta_j(z) \coloneqq \theta_j(\pi z)$ \cite{mckean_moll_1997}, see Section 20.1 of \cite{NIST:DLMF}.} 

We are now ready to analyze the behavior of the two-point functions as $A\to 1$. It is convenient to write $A=1+\delta$ where $0<\delta\ll 1$, which corresponds to weak warp limit. Also note that in this limit, $\Delta \zeta_\pm \xrightarrow[]{}\Delta z_\pm$.

For the Hadamard function (anti-commutator vacuum expectation value), the limit is more transparent in the series form, so we will use Eqs.~(\ref{eq: ci+}). For $C^+_1$ and $C^+_2$ the series expansions are somewhat tedious, but we can see what happens when we take the limit term-wise, since distributionally the series are convergent. Taking the limit term-wise, we see that $\lim_{\delta\to 0} C^+_2(\sx,\sx') = 0$. This follows from the fact that $\delta\to 0$ implies that each term in $C^+_2$ is exponentially suppressed by \mbox{$(\sinh 2\pi^2\beta n)^{-1} \sim 2e^{-n/\delta}$}.

For $C^+_1$, the term-wise limit $\delta\to 0$ gives
\begin{align}
    &\lim_{\delta\to 0}C^+_1(\sx,\sx') \notag\\
    &= \sum_{n=1}^\infty \frac{1}{2 \pi  n} \bigg[\cos \bigg(-\frac{2 \pi  n \Delta z_-}{L}\bigg) +\cos \bigg(\frac{2 \pi  n\Delta z_+}{L}\bigg)\bigg]\notag\\
    &= -\frac{1}{4\pi}\bigg[\log \bigg(1-e^{-\frac{2 \ii \pi  \Delta z_-}{L}}\bigg)+\log \bigg(1-e^{\frac{2 \ii \pi  \Delta z_+}{L}}\bigg)\bigg]\notag\\
    &\hspace{0.4cm}-\frac{1}{4\pi}\bigg[\log \bigg(1-e^{\frac{2 \ii \pi  \Delta z_-}{L}}\bigg) +\log \bigg(1-e^{-\frac{2 \ii \pi  \Delta z_+}{L}}\bigg)\bigg]\notag \\
    &= C^+_{\osc}(\sx,\sx')\,,
    \label{eq: hadamard-EC-oscillator}
\end{align}
where $z_{\pm}$ are the flat space null coordinates. Note that the last equality is precisely the oscillator part of the Hadamard function in the Einstein cylinder calculated in Eq.~(\ref{eq: osc-hadamard}).

Finally, since $\log(1+\delta)\approx \delta$ for $\delta \ll 1$, note that for small $\W\approx \delta/L$, we have
\begin{align}
    \frac{\zeta_+}{\zeta_-} \approx \frac{1+\W z_+}{1-\W z_-}\approx 1+\frac{2\delta}{L}t+ O(\delta^2)\,.
    \label{eq: ratio}
\end{align}
Therefore we have the asymptotic expansion 
\begin{align}
    C^+_0(\sx,\sx')
    &\approx \frac{1}{2\pi\delta}\left[\rr{\frac{2\delta}{L}t' }\rr{\frac{2\delta}{L}t}+\pi^2\right] \notag\\
    &\approx \frac{\pi}{2\delta} + \frac{2\delta^2}{\pi L^2}tt' \,.
    \label{eq: time-machine-zero-mode}
\end{align}
It is clear that $C_0^+$ diverges in the limit $\delta\to 0$. If we set 
\begin{align}
    \gamma\coloneqq 2\delta/\pi\,,
    \label{eq: quantum-harmonic-frequency-parameter}
\end{align}
we obtain for small $\delta$
\begin{align}
    C^+_0(\sx,\sx')  \approx \frac{1}{\gamma} + \frac{\gamma}{2L^2}tt' \,,
\end{align}
which is precisely the Hadamard function of the zero mode in the Einstein cylinder with frequency parameter $\gamma$ in Eq.~(\ref{eq: zm-hadamard}). Therefore, the divergence of $C^+_0$ as $\delta\to 0$ is equivalent to the statement that momentum eigenstate of free particle is completely delocalized in space and has infinite variance ($\Delta Q = \braket{Q^2}-\braket{Q}^2 \to\infty$).

Let us note that $C_0^+$ can be also written in terms of $(\sigma,\chi)$ coordinates. Using Eq.~\eqref{eq: adapted-coordinates}, for $\zeta_\pm>0$  we directly see that, because of Eq.~\eqref{eq: log-sigma}, $C_0^+$  will be just a function of timelike $\sigma$. In the weak warp limit $\delta \to 0$, it can be directly seen  that $\sigma \approx (\delta/L) t$ for small $\delta$, and $C_0^+$ in the adapted coordinates takes the simple expression
\begin{align}
    & C^+_0(\sx,\sx')  \approx  \frac{1}{2\pi\delta }\rr{ 4\sigma\sigma'+\pi^2}\,,
\end{align}
showing that, when restricted to the diamond patch the contribution of $u_0$ to the expectation of the anti-commutator is exactly the same as the contribution of the Einstein-cylinder zero mode.

Let us now compute the Pauli-Jordan function. Using the mode sum formulation for $C^-_\osc(\sx,\sx')$, we get
\begin{align}
    &C^-(\sx,\sx')\notag\\
    &= \frac{\ii\beta}{4}\left[(s_++s_-)\log\left|\frac{\zeta'_+}{\zeta'_-}\right|-(s'_++s'_-)\log\left|\frac{\zeta_+}{\zeta_-}\right|\right] \notag\\
    &\hspace{0.4cm} +\frac{\ii}{2\pi}\sum_{n=1}^\infty \frac{1}{n\sinh( 2\pi^2\beta n)}\notag\\
    &\hspace{0.4cm} \times\bigg[\sinh \Bigr( \pi^2\beta n(s_++s'_+)\Bigr) \sin\rr{2\pi\beta n\log\left|\frac{\zeta_+'}{\zeta_+}\right|}\notag\\
    &\hspace{0.6cm}
    -\sinh \Bigr(\pi^2\beta n(s_-+s'_-)\Bigr) \sin\rr{2\pi\beta n\log\left|\frac{\zeta_-'}{\zeta_-}\right|}\notag\\
    &\hspace{0.6cm} 
    +\sinh \Bigr(\pi^2\beta n(s'_--s_+)\Bigr) \sin\rr{2\pi\beta n\log\left|\frac{\zeta_-'}{\zeta_+}\right|}\notag\\
    &\hspace{0.6cm}
    + \sinh \Bigr(\pi^2\beta n(s_--s'_+)\Bigr)\sin\rr{2\pi\beta n\log\left|\frac{\zeta_+'}{\zeta_-}\right|}\bigg]\,.
\end{align}
For our purposes again we restrict the analysis to the region without CTCs, by setting $s_+=s_-=1$ ($\zeta_\pm>0$), because of our aim of selecting the zero-mode state by obtaining the limit to the Einstein cylinder. The sum can be done analytically, which reads
\begin{align}
    &C^-(\sx,\sx') =  \frac{\ii  \beta}{2}  \left(\log \frac{\zeta _+'}{\zeta _-'} - \log \frac{\zeta _+}{\zeta _-}\right)\notag\\
    &+ \frac{1}{4\pi}\left(\log \left[1-\left(\frac{\zeta _-'}{\zeta _-}\right)^{2 \ii \pi  \beta }\right]-\log \left[1-\left(\frac{\zeta _-'}{\zeta _-}\right)^{-2 \ii \pi  \beta }\right]\right.\notag\\
    &\left.+ \log \left[1-\left(\frac{\zeta _+'}{\zeta _+}\right)^{-2 \ii \pi  \beta }\right]-\log \left[1-\left(\frac{\zeta _+'}{\zeta _+}\right)^{2 \ii \pi  \beta }\right]\right) \,.
\end{align}

Following the same procedure as before, note that the first line is the zeroth mode contribution  which, when restricted to the diamond and using adapted coordinates, is only a function of $\sigma$.  Using the previous small-$\W$ identification $\sigma\approx (\delta /L)t$, we get 
\begin{align}
    C_0^- &\approx \frac{\ii}{ \delta}\rr{-\Delta\sigma}+O(\delta^2) = -\frac{\ii}{L}\Delta t+O(\delta^2)\,,
\end{align}
where $\Delta \sigma = \sigma-\sigma'$ and $\Delta t = t-t'$. Thus in adapted coordinates and in the small $\delta$ limit the connection between $C_0^-$ in the time machine geometry and the zero mode commutator in the Einstein cylinder is manifest: we see explicitly what we already discussed before, i.e., (when restricted to the diamond patch) for all intents and purposes $u_0$ `becomes' the Einstein cylinder zero mode in the $\delta\to 0$ limit.

For small $\delta$, we have $\W\approx \delta/L$ and if we use the identity $e = \lim_{\delta\to 0^+}(1+\delta)^{1/\delta}$, we can show that the limit as $\delta\to 0$ (corresponding to the absence of a time machine) is given by the \textit{full} commutator in the Einstein cylinder with periodicity length $L$ 
\begin{align}
    &\lim_{\delta \to 0} C^-(\sx,\sx') = -\frac{\ii \Delta t}{L} \notag\\
    &
    + \frac{1}{4\pi}\bigg[\log \bigg(1-e^{ -\frac{2 \ii \pi  \Delta z_-}{L}}\bigg)+\log \bigg(1-e^{ \frac{2 \ii \pi  \Delta z_+}{L}}\bigg) \bigg]\notag\\
    &-\frac{1}{4\pi} \bigg[\log\bigg(1-e^{\frac{2 \ii \pi  \Delta z_-}{L}}\bigg)+ \log\bigg(1-e^{-\frac{2 \ii \pi  \Delta z_+}{L}}\bigg)\bigg]\notag\\
    &= C^-_\zm(t,t') + C^-_\osc(\sx,\sx')\,,
    \label{eq: commutator-EC-limit}
\end{align}
where $C^-_{\zm}(t,t') = -\ii\Delta t/L$ and $C^-_{\osc}(\sx,\sx')$ are the remaining terms of the commutator in the Einstein cylinder. The last equality
agrees with the commutators computed in \cite{tjoa2019zeroresponse}. 

Having analyzed in detail both Hadamard and commutator functions, we have determined the behavior of Wightman function. There are few subtleties related to the presence of the zero mode in the limiting case. The fact that in the limit $\delta\to 0$ we recover the zero-mode commutator implies that the zero-mode contribution to the Wightman function is essential for the consistency of the underlying QFT. We cannot simply drop or neglect terms of order $O(\delta^{-1})$ that appear in the Hadamard function $C^+$ by hand, as they are related to the Hadamard function of the zero mode $C^+_\zm$ in the limit $\delta \to 0$. Since one also has to remove the $C^-_\zm$ to neglect the zero mode, this would imply causality violation of the underlying QFT \cite{tjoa2019zeroresponse}. A more conservative attitude would be to restrict attention to only field observables that do not see the zero mode, e.g. shift-invariant operators \cite{Page_2012} or field derivatives \cite{francesco2012conformal}, instead of removing the zero mode from the equations.

Finally, let us comment on  other choices of boundary conditions. In \cite{tjoa2019zeroresponse} it was shown that zero mode also appears when one imposes Neumann boundary condition on two-dimensional massless wave equation. We also saw earlier that the zero-mode contributions to the two-point functions $C^\pm_{\zm}$ for the Einstein cylinder strictly comes from the ``zeroth mode'' of the time-machine case $\bar{u}_0$ as $A\to 1$. It can be checked that taking $A\to 1$ limit also leads to divergences in the Hadamard function $C_0^+(\sx,\sx')$ when other boundary conditions such as Neumann ($\lambda=-\pi/2$) or mixed $(\lambda=-\pi/4)$ are chosen.

\section{Renormalized stress-energy tensor for time-machine model}
\label{sec: RSET}

In view of the connection of the limit of the time-machine model with the zero mode of the Einstein cylinder, we might wonder how the zero-mode contribution to the renormalized stress-energy tensor (RSET) in Einstein cylinder arises from the time machine-model RSET. Since we have the Hadamard function for the time-machine model, the RSET can be computed using various methods such as point splitting \cite{birrell1984quantum}, in addition to using the simplifications that arise from working with the universal covering space \cite{Frolov1991locallystatic}. In order to make this comparison, we recall that in the Einstein cylinder, the Fock vacuum expectation of the renormalized stress-energy tensor has two contributions coming from the zero mode and the oscillator modes.

Let us use the null coordinates $z_{\pm}$ on the Einstein cylinder. The oscillator mode contribution $\braket{^{\osc}T_{\mu\nu}}$ reads~\cite{birrell1984quantum}
\begin{align}
    \braket{^{\osc}T_{--}^{(z)}} &= \braket{^{\osc}T_{++}^{(z)}} = -\frac{\pi}{12 L^2}\,,\label{eq: RSET-uu-cylinder}\\
    \braket{^{\osc}T_{-+}^{(z)}} &= \braket{^{\osc}T_{+-}^{(z)}} = 0\,, 
\end{align}
where the superscript $(z)$ refers to the use of the coordinates $z_{\pm}$. The zero mode contribution is computed in \cite{EMM2014zeromode} giving the result
\begin{align}
    \braket{^{\zm}T_{--}^{(z)}} &= \braket{^{\zm}T_{++}^{(z)}} = \frac{\braket{0_\zm|\hat P^2_{\textsc{s}}|0_\zm}}{4L^2}\,,\label{eq: RSET-uu-zm}\\
    \braket{^{\zm}T_{-+}^{(z)}} &= \braket{^{\zm}T_{+-}^{(z)}} = 0\,.  
\end{align}
Note that while the nonvanishing oscillator components are negative  i.e. $\braket{^{\osc}T_{\mu\nu}}<0$, the zero-mode contribution is manifestly positive for any choice of ``candidate'' zero-mode vacuum state $\ket{0_\zm}$.

For the canonical time machine, the RSET in the region of interest in null $\zeta_{\pm}>0$ coordinates~\cite{Frolov1991locallystatic} is given by
\begin{align}
    \braket{T^{(\zeta)}_{\pm \pm}}&=-\frac{F(\beta)}{\zeta_{\pm}^2}\, , \qquad
    \braket{T^{(\zeta)}_{+-}}=\frac{1}{6\pi (\zeta_{+}+\zeta_{-})^2}\,,  \\
    F(\beta) &= \frac{1}{48\pi} - \frac{\beta}{4\pi} + \frac{\beta^2\pi}{12} - 2\pi\beta^2\sum_{n=1}^\infty\frac{ne^{-4\pi^2\beta n}}{1-e^{-4\pi^2\beta n}}\,.
\end{align}
These results, are consistent with the appearance of the conformal trace anomaly in this spacetime, and, when applied to the Cauchy horizon, have been used to argue for its possible quantum instability {\cite{Frolov1991locallystatic,Sushkov:1995fk,Krasnikov1996stability}.

Our aim is to track the zero-mode contribution to the RSET, so we will trail the small-$\delta$ expansion more carefully. The weak warp limit corresponds to $\W L \approx \delta \ll 1$, hence
\begin{align}
    \zeta_- &= e^{\W y}/\W- t  \approx \W ^{-1} + z_-\,,\\
    \zeta_+ &= e^{\W y}/\W+ t \approx \W ^{-1}+ z_+\,.
\end{align}
Consequently, we have that the non-diagonal element of the RSET yields
\begin{align}
    \braket{T^{(z)}_{-+}} &= \frac{\partial \zeta_-}{\partial z_-}\frac{\partial
    \zeta_+}{\partial z_+}\braket{T^{(\zeta)}_{-+}} \notag\\
    &\approx \frac{1}{6\pi(\zeta_{+}+\zeta_{-})^2}=\frac{\delta^2}{24\pi L^2}\,,
\end{align}
which in the limit $\delta\to 0$ gives a vanishing contribution $ \braket{T^{(z)}_{-+}} = 0$. 

Now we focus on calculating the small-$\delta$ limit of the diagonal element of the RSET. They are given by
\begin{align}
    \braket{T_{--}^{(z)}} &= \frac{\partial \zeta_-}{\partial z_-}\frac{\partial
    \zeta_-}{\partial z_-}\braket{T_{--}^{(\zeta)}} \notag\\
    &\approx -\frac{\delta^2}{L^2}\rr{\frac{1}{48 \pi }-\frac{1}{4 \pi  \delta}+ \frac{\pi }{12 \delta ^2}}\notag\\
    &= \frac{\delta}{4 \pi  L^2}-\frac{\pi }{12 L^2} + O(\delta^2)\,,\\
    \braket{T_{++}^{(z)}} &=  \braket{T_{--}^{(z)}} \notag\\
    &\approx \frac{\delta}{4 \pi  L^2}-\frac{\pi }{12 L^2} + O(\delta^2)\,.
\end{align}

Next, using $\beta\approx \delta^{-1}(1+\delta/2)$, we obtain
\begin{align}
    F(\beta) &\approx -\frac{1}{4 \pi  \delta}+ \frac{\pi }{12 \delta ^2}+ O(\delta^0)\,.
\end{align}
An important step here is that we cannot keep only the $O(\delta^{-2})$ term (as done in \cite{Frolov1991locallystatic}), as we will see that the relationship with the zero mode of the Einstein cylinder is of order $O(\delta)$ in the RSET. This corresponds to keeping the $O(\delta^{-1})$ term in the asymptotic expansion for $F(\beta)$. We also keep the $O(\delta^0)$ term in $F(\beta)$ for clarity in what follows.

Observe that $\braket{T_{--}^{(z)}}$ and $\braket{T_{++}^{(z)}}$ contain two contributions, one giving the expected Casimir contribution from the oscillator modes of the Einstein cylinder in Eq.~\eqref{eq: RSET-uu-cylinder}, and another term that is linear in $\delta$. Recall from previous section that for small $\delta$ we identified $\delta$ with the frequency parameter $\gamma= 2\delta/\pi$. With this identification, we can now write
\begin{align}
    \braket{T_{--}^{(z)}} &\approx \frac{1}{4L^2}\frac{\gamma}{2} + \braket{^{\osc}T_{--}^{(z)}}  \,.
\end{align}
However, if the zero mode vacuum $\ket{0_\zm}$ is to be identified with the ground state of quantum harmonic oscillator with frequency parameter $\gamma$, we recall that the second moment of $P_\textsc{s}$ is given by $\braket{P^2_\textsc{s}} = \gamma/2$. Using this, it is now clear that for very small $\delta$ the RSET can be written as
\begin{align}
    \braket{T_{--}^{(z)}} &\approx \braket{^{\zm}T_{--}^{(z)}} + \braket{^{\osc}T_{--}^{(z)}}  \,.
\end{align}
The fact that $\braket{^{\zm}T_{--}^{(z)}}$ is $O(\delta)$ shows that indeed as $\delta\to 0$, the zero mode of the time machine picks out the eigenstate of $P$ of quantum mechanical free particle with zero momentum eigenvalue (i.e. $\gamma=0$). This is consistent with the performed computation \mbox{$\braket{^{\zm}T_{--}^{(z)}}\propto \braket{0_\zm|P^2_S|0_\zm}$} which vanishes when $\ket{0_\zm}$ is taken to be the eigenstate of $P$ with zero eigenvalue. Notice that if we were to only keep $O(\delta^{-2})$ term for the asymptotic expansion of $F(\beta)$, we would only recover the contribution from the oscillator modes in the Einstein cylinder and we would not see how the zero-mode contribution appears in the RSET (which happens to vanish for $\gamma=0$).  

The calculations in this section show, on one hand, that the zero mode of a massless scalar field in the Einstein cylinder is connected with the mode $\overline u_0$ in the automorphic solution [Eq.~\eqref{eq: u0}] of the field obtained via the universal covering construction. 
This connection is not manifest when we only keep the term that grows as $\delta^{-2}$ in $F(\beta)$ and take the $\delta\to 0$ limit.  
On the other hand, since all the modes of the time-machine model have a Fock representation, we can think of the time-machine model at small $\delta$ as being a small deformation from the geometry of the Einstein cylinder 
that does not have a zero mode. These results provide then a natural way to remove the zero mode ambiguity, by fixing the quantization in the Einstein cylinder from its deformations. 

Another possibility we might have considered is the creation of a time machine at $t=0$, instead of having an eternal time machine. In this case one begins (at $t=-\infty$) by having an Einstein cylinder spacetime that matches the time-machine metric at $t=0$. Then the Poincar\'e patch is defined as in our case but restricting it to the domain of $\eta >0$. Note that in this case the spacetime possesses only a bifurcate future Cauchy horizon, given by $\mathcal{H}^+$ and $\mathcal{H'}^+$ in the conformal diagram in Figure~\ref{fig: penrose}.

\section{Conclusion}

Inspired by the study of time machines, we propose a way to solve the inherent ambiguity of the zero-mode quantization in QFT living in spacetimes with spatial periodicity (such as the Einstein cylinder).  The appearance of zero modes in QFT has traditionally been ignored in many QFT calculations (see e.g. \cite{birrell1984quantum,Lin2016entangleCylin,Robles2017thermometryQFT,Brenna2016antiUnruh,braun2005entanglement,Lorek2014tripartite}), but it has been recently shown to be necessary in order to conserve the relativistic aspects of the measurable predictions of the theory~\cite{tjoa2019zeroresponse}.  The main problem with the zero mode in a spatially periodic spacetime is that there is no good reason to select a state and declare it {\it the vacuum} since the zero mode does not admit a Fock quantization as it is dynamically equivalent to quantum mechanical free particle (see, e.g.,~\cite{EMM2014zeromode,tjoa2019zeroresponse,tjoa2020zeroharvest}). 

Concretely, we have studied the quantization of a scalar field in a spacetime with time machines. In particular we have considered a time-machine model, that corresponds to the $(1+1)$-dimensional locally static multiply-connected spacetime first studied in~\cite{Frolov1991locallystatic}. To get a rough idea about this spacetime, one can think of an Einstein cylinder (a spacetime with a spatial periodicity in a particular time foliation, akin to imposing periodic boundary conditions)  in which the points that are topologically identified by the periodicity are in different time slices. Because of that, this spacetime can contain closed timelike curves in some region. The analysis in the subregion without CTCs shows that in the limit that there is no time shift between the topologically identified points one recovers the quantization on the Einstein cylinder, including the zero-mode contribution.

We have compared the quantization of a massless scalar field in this time-machine spacetime with the quantization in an Einstein cylinder (which, we recall, corresponds to a spacetime with periodic boundary conditions). Classically, the Einstein cylinder is obtained from the limit where there is no time machine in the locally static model spacetime. However, when we consider a quantum field in the time-machine model spacetime, there is no zero-mode ambiguity. Crucially, we find that quantization of a massless scalar field in a time-machine spacetime induces a unique quantization of a massless scalar field with periodic boundary conditions, with the zero mode of the Einstein cylinder appearing naturally in the weak warp limit.
We obtain the zero mode by explicitly tracking the appearance of a time-warp parameter $A\geq 1$ that performs the time identification $t\sim At$ in the field modes in the time-machine model.

From our results, we can construct the following prescription for a zero-mode quantization in spatially periodic spacetimes. 1) Introduce a deformation parameter in the Einstein cylinder by adding some small time shift in  addition to the spatial periodicity of the spacetime, i.e. consider the one-parameter family of time-machine spacetime with warp parameter $A\geq 1$; 2) Perform a quantization of the field in the universal covering spacetime of this one-parameter family of geometries, which removes zero modes and all its ambiguities; 3) Take the limit of the warp parameter $A\to 1$ where the spacetime becomes the Einstein cylinder again. The zero mode emerges from the limit and a state is selected. This construction  supports the choice made for simplicity {(i.e. as a squeezed state)} in e.g. in \cite{EMM2014zeromode, tjoa2019zeroresponse} for state for the zero mode to evaluate its impact on particle detector dynamics.
 Thus, we propose that there is a unique way of selecting the state of the zero mode in a periodic spacetime that is compatible with the quantization in the universal covering. 

Furthermore, we argue that the vacuum of the scalar field in the time-machine spacetime proposed in~\cite{Frolov1991locallystatic} yields a divergent Hadamard function for the emergent zero mode of the Einstein cylinder when we consider the no-time-machine limit. This shows that while the classical background metric for the time machine has a smooth limit to the Einstein cylinder, the quantum field theory on the time-machine model does not smoothly reduce to the Einstein cylinder limit due to divergence in the {\it zero-mode component} of the two-point Wightman functions. The emergent zero-mode state is akin to an (unnormalizable) eigenstate of a momentum operator. This is easily solvable by regularizing that state as a {\it squeezed} state, and we propose to do this following on previous results in~\cite{EMM2014zeromode, tjoa2020zeroharvest}, or equivalently by regularizing the Einstein cylinder quantization using small warp parameter (very weak time-machine geometry).

Finally, we have analyzed the stress-energy tensor both in the time-machine spacetime and in the Einstein cylinder with this regularized zero-mode state, showing that the behavior is regular as expected. These results shed light on quantum field theory in the presence of time machines, and can be used to resolve zero-mode ambiguities in QFT in spacetimes with periodic boundary conditions.

\begin{acknowledgments}
A. A-S. is supported by the ERC Advanced Grant No. 740209. E.T. acknowledges support from Mike and Ophelia Lazaridis Fellowship. E.M.-M. acknowledges support through the Discovery
Grant Program of the Natural Sciences and Engineering Research Council of Canada (NSERC). E.M.-M. also acknowledges support of his Ontario Early Researcher award. This work is funded by  Project No. MICINN PID2020-118159GB-C44 from Spain.
\end{acknowledgments}

\appendix

\section{Four-acceleration and four-velocity of Killing observers}
\label{appendix: four-accel}

In this appendix we show that the condition that there exists a hypersurface-orthogonal Killing field $\xi^\mu$, i.e. that satisfies
\begin{align}
\nabla_{(\mu}\xi_{\nu)}=0, \, \quad \xi_{[\lambda} \nabla_{\mu}\xi_{\nu]}=0
\label{eq: killing-field}
\end{align}
is equivalent to the existence of a timelike vector field $u^\mu$ and a spacelike vector field $a^\mu$ such that
\begin{align}
u^\mu u_\mu=-1, \,\quad \nabla_\nu u_\mu=-a_\mu u_\nu, \, \quad  \nabla_{[\mu}a_{\nu ]}=0.
\label{eq: killing-observers}
\end{align}
Let us first show that Eq.~(\ref{eq: killing-observers}) implies Eq.~(\ref{eq: killing-field}). 

From $\nabla_{[\mu}a_{\nu]}=0$ we conclude that there exists some $\varphi$ such that
\begin{align}
    a_\mu=\nabla_\mu \varphi.
    \label{eq: acceleration}
\end{align}
Let us check that the vector field  $\xi^\mu=e^\varphi u^\mu$ satisfies Eq.~(\ref{eq: killing-field}). Indeed, using Eq.~(\ref{eq: killing-observers}) and Eq.~(\ref{eq: acceleration}), we see that
\begin{align}
    \nabla_{(\mu}\xi_{\nu)}&=e^\varphi \nabla_{(\mu}u_{\nu)}+e^\varphi \nabla_{(\mu}\varphi u_{\nu)} \nonumber \\
    &=-e^\varphi a_{(\nu}u_{\mu)} + e^\varphi a_{(\mu}u_{\nu)}=0\,.
\end{align}
On the other hand,
\begin{align}
\xi_{[\lambda}\nabla_\nu \xi_{\mu]}=e^\varphi u_{[\lambda}\nabla_{\nu}e^\varphi u_{\mu]}+e^{2\varphi}u_{[\lambda}\nabla_\nu u_{\mu]}.
\end{align}
The first term trivially vanishes. The second one vanishes as well because $\nabla_\nu u_\mu=-a_\mu u_\nu$. 

Let us now prove that Eq.~(\ref{eq: killing-field}) implies Eq.~(\ref{eq: killing-observers}). Given $\xi^\mu$ satisfying Eq.~(\ref{eq: killing-field}), let us define $u^\mu=e^{-\varphi}\xi^\mu$ with $\varphi=\log |\xi|$. The field $u$ obviously satisfies $u^2=-1$. If we define $a_\mu=u^\nu \nabla_\nu u_\mu$, then it is straightforward to see that
\begin{align}
    a_\mu=e^{-2\varphi}\xi^\nu \nabla_\nu \xi_\mu-e^{-2\varphi}\xi_\mu \xi^\nu \nabla_\nu \varphi.
    \label{eq: four-accel-intermediate}
\end{align}
Using $\nabla_{(\mu}\xi_{\nu)}=0$ it is easy to see that the first term is
\begin{align}
    -e^{-2\varphi}\xi^\nu \nabla_\mu \xi_\nu = -\frac{1}{2}e^{-2\varphi} \nabla_\mu |\xi|^2=\nabla_\mu \varphi\,,
\end{align}
where in the last equality we have used $|\xi|^2 = -e^{2\varphi}$. Furthermore,  the second term of Eq.~\eqref{eq: four-accel-intermediate} is proportional to
\begin{align}
    e^{-2\varphi}\xi^\nu \nabla_\nu \varphi=-\xi^\nu \nabla_\nu(\xi^\rho \xi_\rho)= -2\xi^\nu\xi^\rho\nabla_\nu\xi_\rho =0,
\end{align}
where  we have used the fact that $\nabla_{(\mu}\xi_{\nu)}=0$ (hence $\nabla_{\nu}\xi_\rho$ is antisymmetric). Therefore, we conclude that $a_\mu=\nabla_\mu \varphi$ and consequently $\nabla_{[\mu}a_{\nu]}=0$. 

Finally, let us check that the following  quantity vanishes:
\begin{align}
\nabla_\mu u_\nu +a_\nu u_\mu&=e^{-3\varphi}(\xi^\rho \xi_\nu \nabla_\mu \xi_\rho +\xi^\rho \xi_\mu \nabla_\rho \xi_\nu) \nonumber \\
& +e^{-\varphi}\nabla_\mu \xi_\nu +e^{-5\varphi}\xi_\mu \xi^\lambda \xi^\rho \nabla_\lambda \xi_\rho \xi_\nu.
\end{align}
Using Eq.~(\ref{eq: killing-field}) we see, on one hand, that the first term cancels the second one. On the other hand, the last term also vanishes because $\nabla_{\lambda}\xi_{\rho}$ is antisymmetric.

\section{Multiply-connected spacetimes and automorphic functions}
\label{appendix: automorphic}

Here we outline the basic idea of automorphic forms in the context of quantum field theory in multiply-connected spacetimes. We will use the notation consistent with the main text: we denote by $\bar{M}$ the multiply-connected space and $M$ its universal covering space that is simply connected. In particular, in this paper we have $M=\R^2$ and $\bar{M}=\R\times S^1$.

We say that a manifold is \textit{simply connected} if it is path-connected (i.e. any two points can be connected by a continuous curve) and any closed loop can be continuously deformed to a single point. Otherwise we say that it is \textit{multiply connected}. For example, the unit circle $S^1$ is not simply connected while higher-dimensional spheres $S^n$ ($n\geq 2$) are simply connected. 

The \textit{fundamental group} $\pi_1(\bar{M})$ of a multiply-connected manifold $\bar{M}$ characterizes the different possible closed loops on the manifold up to continuous deformation. For simply connected manifolds, there is essentially only one closed loop---equivalent to a single point---since we can always ``shrink'' them continuously. Therefore the fundamental group of a simply connected $M$ consists of only a single element, the identity $e$, i.e. $\pi_1( M) = \{e\}$. In contrast, $S^1$ is not simply connected because we cannot continuously shrink any loop to a point, and loops that go through the circle different number of times are not equivalent. Hence we write $\pi_1(S^1) \cong \mathbb{Z}$. Note that if $\bar M$ can be written as a product $\bar{M} \cong Y\times Z$ where $Y$ is simply connected and $Z$ is multiply connected, then $\pi_1(\bar{M})$ is isomorphic to $\pi_1(Z)$.

The \textit{universal covering space} $M$ of a multiply-connected manifold $\bar M$ is obtained by ``unwrapping'' $\bar M$. The universal covering space of $\bar M$ is essentially unique and is simply connected. In general, the multiply-connected space $\bar M$ is always related to its universal covering by the (discrete) group action of the fundamental group. In the language of group theory, points in the quotient space $\bar M$ are \textit{equivalence classes} of points in $M$ given by the equivalence relation $x\sim g\cdot x$ where $g\in \pi_1(\bar M)$, often denoted by $[x]\in \bar M$. In other words, every point $[x]\in \bar M$ is the orbit $[x]=\{g\cdot x: g\in \pi_1(\bar M)\}$ and $\bar M$ is the set of orbits of the fundamental-group action, which we write as $\pi_1(\bar M) \cdot M$ (as a quotient space this is sometimes written as $\pi_1(\bar M)\!\setminus\! M$.)
 
A \textit{fundamental domain} of the fundamental-group action $\pi_1(\bar M)\cdot  M$ is a connected open subset $U$ of $ M$ with the property that the collection $X= \{g\cdot U:g\in \pi_1(\bar M)\}$ is disjoint and the closure $\overline{X}$ covers $ M$. Thus for every fundamental domain $U$, the set $g\cdot U$ (for fixed $g$ and $U$) contains exactly one point from each orbit $[x]$ for every $x\in U$. The fact that $[x]$ is an orbit means that in practice one can identify the multiply-connected manifold $\bar{M}$ itself with a fundamental domain.

A function $f: {M} \to \R$ is a  \textit{global automorphic form} on ${M}$ with respect to a symmetry group $\Gamma$ (in our case $\Gamma=\pi_1(\bar M)$) if for $x\in  M$, $g\in \Gamma$, we have
\begin{align}
    f(g\cdot x) = a(g)f(x)\,,
    \label{eq: automorphy}
\end{align}
where $a(g)$ is a constant called \textit{automorphic factor}. The automorphic factor has the property that $a(g)= 1$ if and only if $g=e$ (the identity element of $\Gamma$) and $a(g_1g_2)=a(g_1)a(g_2)$.

Let us now proceed with the analysis of  real scalar fields $\phi$ on $M$ and $\bar\phi$ on $\bar M$, whose dynamics is given by the following Lagrangian density
\begin{align}
    L[\phi(x)]\coloneqq -\frac{1}{2}\nabla_\mu\phi\nabla^\mu\phi - \frac{1}{2}m^2\phi^2 - \frac{1}{2}\xi R\phi^2\,.
\end{align}
Since the Lagrangian density is defined locally, the same expression applies for $\bar\phi$. On ${M}$, we do not need $\phi$ to have any additional symmetry. However, if we are to study scalar field theory on $\bar M$ by going into its universal cover $M$, the relationship between $\phi$ and $\bar\phi$ implies that the Lagrangian density $L[\phi]$ should obey additional symmetry requirements imposed by $\Gamma$. The requirement that the Lagrangian is the same for {both fields} means that $\phi$ respects the action of $\Gamma$. In other words, $\phi$ should be an automorphic form,
\begin{align}
     \phi(g\cdot x) &= a(g)\phi(x)\,.
\end{align}
Furthermore, since $\bar\phi$ depends only on orbits $[x]$, the action of $\Gamma$ on $M$ should be regarded as a symmetry operation and hence the Lagrangian should be invariant under $\Gamma$, i.e. 
\begin{align}
    g\cdot L[\phi(x)]\coloneqq  L[\phi(g\cdot x)] = L[\phi(x)]\,.
\end{align}
Since the Lagrangian $L[\phi]$ is quadratic and that $\nabla_\mu\phi(g\cdot x) = a(g)\nabla_\mu \phi(x)$, it means that the automorphic factor satisfies $a(g)^2=1$, and for real-valued $\phi$ it means that $a(g)=\pm 1$.

The scalar field $\bar \phi$ can now be obtained from $\phi$ via the following averaging procedure
\begin{align}
    \bar\phi([x]) &= \frac{1}{|\Gamma|}\sum_{g\in\Gamma} a(g^{-1})\phi( g\cdot x)\,,
    \label{eq: field-averaging}
\end{align}
where strictly speaking the averaging should be considered carefully since in our case $\Gamma=\pi_1(\bar{M})\cong \mathbb{Z}$ and hence $|\Gamma|=\infty$ \cite{Banach1979mathissues,Banach1980automorphic,BanachDowker1979}. The requirement that $\phi$ is an automorphic form can be expressed as
\begin{align}
    \bar{\phi}([x]) &= \frac{1}{|\Gamma|}\sum_{\gamma\in \Gamma}a(g^{-1})a(g)\phi(x) = \phi(x)\,.
\end{align}
In other words, $\phi$ is automorphic if and only if $\phi(x)=\bar\phi([x])$. This seemingly ``trivial'' result says that at the level of physical calculations, we do not need to distinguish $\phi$ from $\bar\phi$ apart from the recognition that $\bar\phi$ is a function of orbits $[x]\in \bar M$. In other words, the fields on $\bar M$ are essentially the same as the fields on $M$ but with ``generalized periodic boundary conditions'' on $\bar\phi$.

A useful result from this construction is that we can relate the Wightman two-point functions in $\bar M$ and $ M$ for scalar fields. In particular, we have \cite{Banach1979mathissues}
\begin{align}
    W_{\bar M}(\sx,\sx') = \sum_{g\in\Gamma}W_{M}(\sx,g\cdot \sx')a(g)\,.
    \label{eq: automorphic-Wightman}
\end{align}
This expression is very convenient because it does not involve averaging even if the cardinality of $\Gamma$ is infinite.

Quantum field theory on the universal covering space is typically straightforward, since objects like global timelike Killing vector fields are often available for the definition of positive- and negative-frequency solutions. If $\Sigma$ is a Cauchy surface on $ M$ which is invariant under the action of $\Gamma$, then we can define the Klein-Gordon inner product as usual
\begin{align}
    \braket{\phi_1,\phi_2} = -\ii\int_{\Sigma} \d\Sigma^\mu\, [\phi_1\partial_{\mu} \phi_2^*-\phi_2^*\partial_{\mu} \phi_1]\,.
\end{align}
The fact that the automorphic condition imposes $\phi=\bar\phi$ means that if we take $\bar\Sigma={\Sigma}\cap \bar M$, then the corresponding induced Klein-Gordon inner product on $\bar M$ reads
\begin{align}
    (\bar\phi_1,\bar\phi_2) = -\ii\int_{\bar\Sigma} \d\Sigma^\mu\, [\bar\phi_1\partial_{\mu} \bar\phi_2^*-\bar\phi_2^*\partial_{\mu} \bar\phi_1]\,,
    \label{eq: KG-inner-product-induced}
\end{align}
which is just the inner product under the restriction of $\Sigma$ to the fundamental domain $\bar M$ and $\phi$ necessarily automorphic. 

\section{Boundary conditions}
\label{appendix: boundary-conditions}

In order to better understand the possible boundary conditions that are compatible on the Poincar\'e patch, it is instructive to first consider $\phi$ to be a real, massive scalar field with an arbitrarily coupling to curvature, whose equation of motion reads 
\begin{align}
    \rr{-\partial_\eta^2+\partial_\xi^2 - \frac{m^2}{\W^2\xi^2}}\phi(\eta,\xi) = 0\,,
\end{align}
where $m^2 = m_0^2 + \kappa R$ is the effective mass that depends on the field's bare mass $m_0$ and Ricci scalar $R$, with $\kappa$ being an arbitrary  constant. The field $\phi$ can be written in terms of its Fourier transform
\begin{align}
    \phi(\eta,\xi) = \int \d\omega \,e^{-\ii \omega \eta}\Phi_\omega(\xi)\,,
\end{align}
where {the modes} $\Phi_\omega(\xi)$ satisfy the second-order ordinary differential equation
\begin{align}
    \xi^2\frac{\d^2\Phi_\omega}{\d \xi^2}  + \rr{\omega^2\xi^2-\frac{m^2}{\W^2}}\Phi_\omega = 0\,.
    \label{eq: ODE-Poincare}
\end{align}
The general solution of this equation is given by linear superposition of two linearly independent solutions, namely
\begin{align}
    \Phi_\omega(\xi) = c^+ (\omega) \Phi_{\omega}^+(\xi)+c^- (\omega) \Phi^-_{\omega}(\xi)\,.
    \label{eq: ODE-solution-Poincare}
\end{align}
Here $c^\pm (\omega)$ are constants and the two fundamental solutions $\Phi^\pm_\omega(\xi)$ are given by\footnote{$J_{\pm\nu}$ are linearly independent so long as $\nu\not\in \mathbb{Z}$. When $\nu\in \mathbb{Z}$, we take $\Phi^+_\omega =\sqrt{\W\xi} J_\nu(\omega\xi)$ and $\Phi^-_\omega = \sqrt{\W\xi} Y_\nu(\omega\xi)$, where $Y_\nu(\omega\xi)$ is the Bessel function of the second kind.}
\begin{align}
    \Phi_{\omega}^\pm (\xi) = \sqrt{\W\xi}J_{\pm\nu}( \omega \xi)\,,
\end{align}
where $J_{\pm\nu}$ are Bessel functions of the first kind~\cite{abramowitz1972handbook} and \mbox{$\nu = \frac{1}{2}\sqrt{1+4m^2\W^{-2}}$}. We will consider $m^2\W^{-2}\geq -\frac{1}{4}$ so that $\nu\in [0,\infty)$. This lower bound on $m^2$ is known as the Breitenlohner-Freedman bound \cite{Breitenlohner1982bound}. This bound is analogous to demanding that a quartic oscillator with anharmonic potential $g x^4$ should have $g\geq 0$ for a stable ground state to exist.

The most general boundary conditions on $\phi(\eta,\xi)$ at $\xi=0$ can be summarized as a one-parameter family of Robin boundary conditions~\cite{Dappiaggi2016PAdS}:
\begin{align}
    \cos(\lambda)\phi(\eta,\xi)\Bigr|_{\xi=0} +  \sin(\lambda)\frac{1}{\W}\frac{\d}{\d \xi}\phi(\eta,\xi)\Bigr|_{\xi=0} =0\,,
    \label{eq: Robin-boundary conditions}
\end{align}
with $\lambda\in [-\pi/2,0]$, or equivalently the one-parameter family of boundary conditions on the Fourier transform modes $\Phi_\omega(\xi)$:
\begin{equation}
    \bc(\xi):=(\cos\lambda)\Phi_\omega(\xi)+\W^{-1}(\sin\lambda) \Phi_\omega'(\xi)=0.
\label{eq:spectralbc}
\end{equation}
The Dirichlet boundary condition ($\lambda=0$) was chosen in \cite{Frolov1991locallystatic} based on the argument that any regular solutions (solutions that do not diverge anywhere) of massive Klein-Gordon equation for arbitrarily low mass on this patch must vanish on the timelike asymptotic boundary $\xi=0$. 

In order to see how the Dirichlet boundary was chosen, we need to study the behavior of the Bessel functions for  small arguments. Substituting Eq. \eqref{eq: ODE-solution-Poincare} into Eq.~\eqref{eq:spectralbc}, we obtain the following behavior close to the conformal boundary $\xi=0$:
\begin{align}
\bc(\xi)&=c^+ (\omega)\bc^+(\xi)+c^- (\omega)\bc^-(\xi),\nonumber
\\ 
\bc^\pm(\xi)&\sim  (\W \xi)^{\frac12\pm\nu}\cos\lambda+(\W \xi)^{-\frac12\pm\nu}(1/2\pm\nu)\sin\lambda,
\label{eq: BCpm}
\end{align}
up to some irrelevant constant prefactor. Therefore, the situation splits into three cases: 
\begin{enumerate}[leftmargin=*,label=(\roman*)]
    \item $m^2/\W^2>0$ ($\nu>1/2$): near $\xi=0$, we have $\text{BC}_\omega^+\sim 0$. However, $\text{BC}_\omega^-\sim A(\W \xi)^{\frac12-\nu}\cos\lambda +  B(\W \xi)^{-\frac12-\nu}\sin\lambda$ which diverges for any $\lambda$ unless $c_-(\omega)=0$. In other words, the field only has contribution from $\Phi_\omega^+$ and hence vanishes at  $\xi=0$, equivalent to taking Dirichlet boundary condition. One heuristic reason why the $m^2>0$ case is only consistent with Dirichlet boundary condition for $\Phi_\omega$ to be regular near the conformal boundary $\xi=0$ is that timelike geodesics cannot reach $\xi=0$ due to refocusing properties of AdS$_2$ \cite{hawking1973largescale}.
     
    \item $-1/4<m^2/\W^2<0$ ($\nu<1/2$): both $\bc^\pm(\xi)$ diverge unless $\lambda =0$. Therefore, the only nontrivial boundary condition that we can impose is the Dirichlet boundary condition.
    
    \item $m^2/\W^2=0$ ($\nu=1/2$): in this case, $\bc^+(\xi)\sim \sin\lambda$ and $\bc^-(\xi)\sim\cos\lambda$. Consequently, all possible boundary conditions---Dirichlet, Neumann, and Robin---can be imposed by suitably fixing $c_\pm(\omega)$.
\end{enumerate}

\bibliography{references}

\end{document}